\documentclass[sigchi]{acmart}

\usepackage{booktabs} 
\usepackage{subcaption}

\graphicspath{{figures/}{pictures/}{images/}{imgs/}{./}}

\setcopyright{none}

\acmConference[]{}{}{}
\acmYear{2018}

\renewcommand\footnotetextcopyrightpermission[1]{}
\begin{document}
\title{Ethical Dimensions of Visualization Research}

\author{Michael Correll}
\orcid{0001-7902-3907}
\affiliation{%
  \institution{Tableau Research}
  \city{Seattle}
  \state{WA}
}
\email{mcorrell@tableau.com}


\begin{abstract}
Visualizations have a potentially enormous influence on how data are used to make decisions across all areas of human endeavor. However, it is not clear how this power connects to ethical duties: what obligations do we have when it comes to visualizations and visual analytics systems, beyond our duties as scientists and engineers? Drawing on historical and contemporary examples, I address the moral components of the design and use of visualizations, identify some ongoing areas of visualization research with ethical dilemmas, and propose a set of additional moral obligations that we have as designers, builders, and researchers of visualizations.
\end{abstract}

%
%



\begin{teaserfigure}
  \centering
  \includegraphics[width=.8\textwidth]{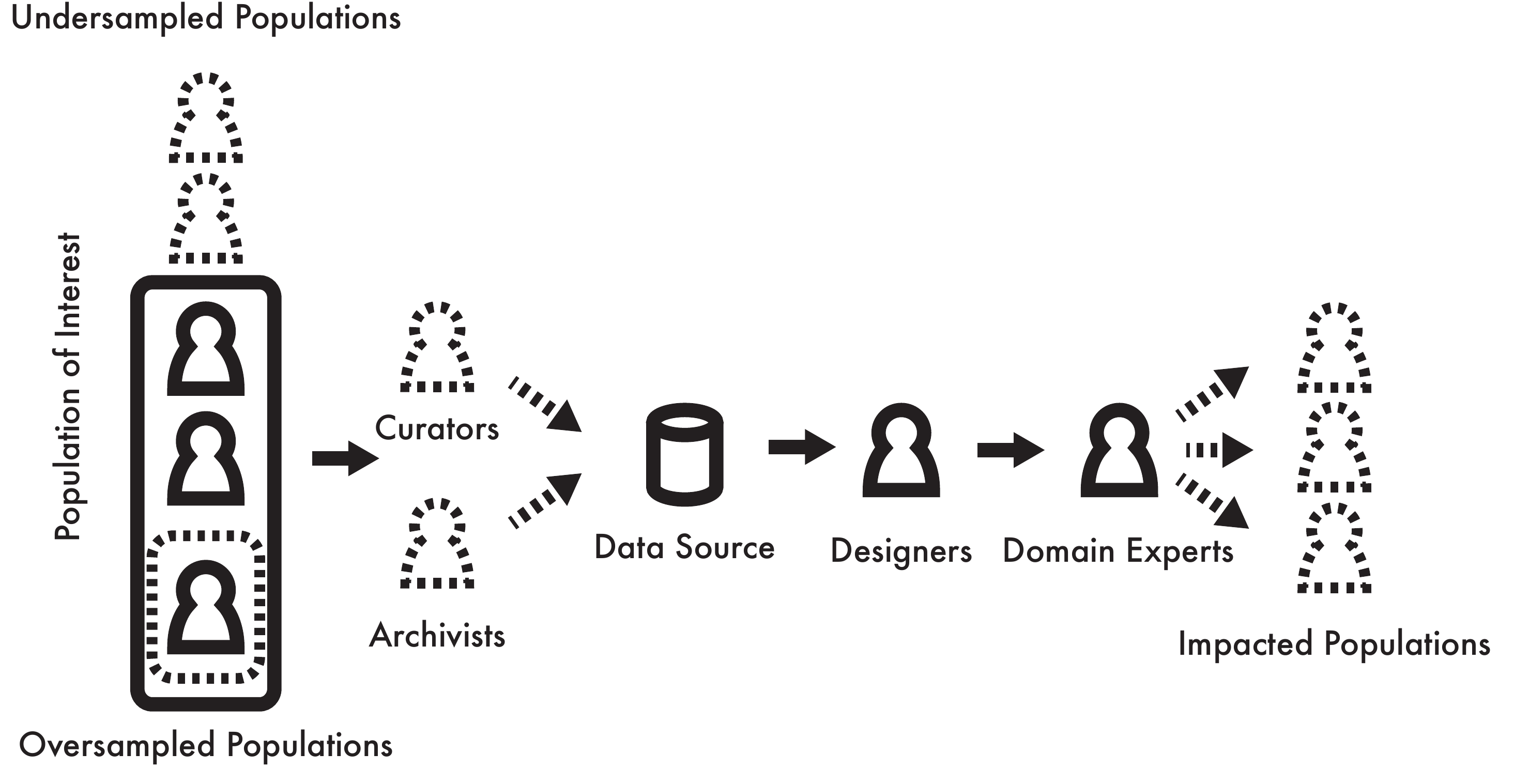}
  \caption{Visualizations projects are often described or evaluated as though they are straightforward paths from data collection to design to the intended user (solid outlines). This neglects or makes invisible critical populations, impacts, and labor (dashed outlines) that can contribute to the ethical character of a project. We have an obligation, where possible, to make these invisible facets and contributions visible.}
  \label{fig:teaser}
\end{teaserfigure}

\maketitle

\section{Introduction}
In the wake of leaked information about the NSA's spying program, Rogaway wrote ``The Moral Character of Cryptographic Work''~\cite{rogaway2015moral}. In that paper, he argues that the work of academics and engineers in cryptography has an inescapable moral character: it shifts power amongst social groups, and so has an inherent political impact on society, for good or ill. Critical movements in cartography~\cite{crampton2006introduction,crampton2011mapping,wood2010rethinking} and data science~\cite{boyd2012critical,dalton2014does} have begun to analyze how the the use (and abuse) of data shifts structures of power. Visualization work has the same capacity, and so must also be analyzed with respect to its moral character. As per Rogaway: 

\begin{quote}
I suspect that many of you see no real connection between social, political, and ethical values and what you work on. You don't build bombs, experiment on people, or destroy the environment. You don't spy on populations. You hack math and write papers. This doesn't sound ethically laden. I want to show you that it is.
\end{quote}

In this paper I will draw on the history of visualization and analytics to illustrate that \emph{all} visualization research, no matter how superficially apolitical or trivial, has a moral character. I will then illustrate how this moral character is reflected in conflicts between virtues that arise in current emerging areas of visualization research, and how they might be balanced. My goal with this work is both to promote caution and contemplation in visualization research (in that we should stop doing unethical work) but also to present new opportunities for research and growth (in that we should study the broader impact of our work and look for new areas to explore and problems to solve).

In the first two sections of this paper I will address the common feeling that data and data visualization, respectively, are apolitical or somehow ethically neutral, and that therefore we lack moral obligations with regards to how data are collected and visualized. It is this tendency to view our work as the mere reporting or structuring of objective fact that is most dangerous to me. Heidegger~\cite{heidegger1954question} specifically calls out the danger of this perspective:
\begin{quote}
Everywhere we remain unfree and chained to technology, whether we passionately affirm or deny it. But we are delivered over to it in the worst possible way when we regard it as something neutral.
\end{quote}

In the final two sections of the paper, I will address current trends in visualization research that appear to have ethical implications. I will then use these case studies as a basis to propose additional obligations that visualization researchers have in addition to their existing moral obligations as scientists, teachers, and citizens.

\section{Against the Neutrality of Data}
\begin{figure}
\centering
\includegraphics[width=.8\columnwidth]{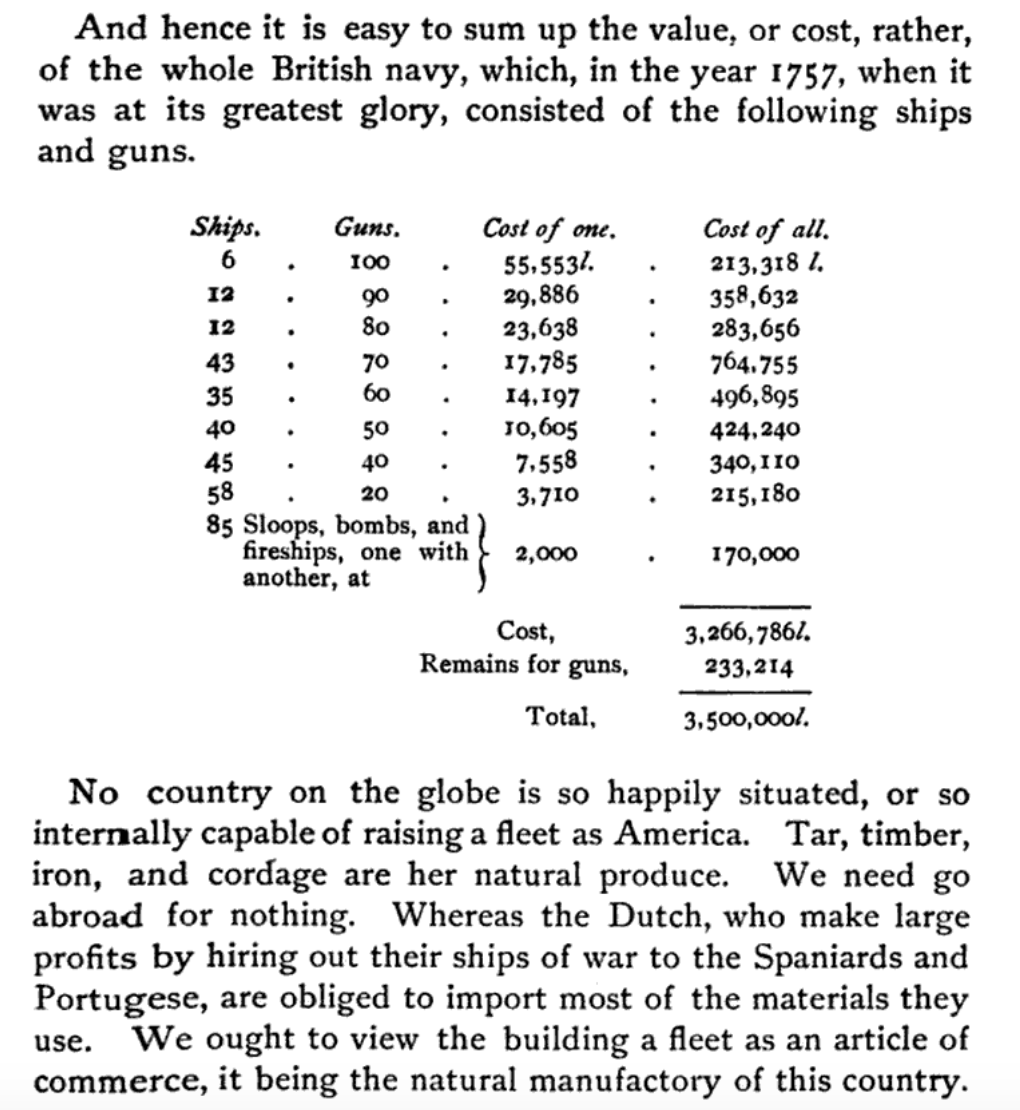}
\caption{A table from Thomas Paine's 1775 pamphlet \emph{Common Sense}\protect\cite{paine2004common}. These data were collected for an initial accounting purpose, but are used by Paine to argue for the relative weakness and fragility of the English Navy, and the potential strength of the American Navy, as part of an argument for independence and revolution.}
\label{fig:paine}
\end{figure}

It is tempting to claim that visualization is an ethically neutral activity because we are merely reporting the data, and data are just facts about the world. It's not our problem how these facts are collected, or who uses them. We're just the middle-man (or, more nefariously, the man in the middle~\cite{correll2017black}) between a stakeholder and their data. Provided that we did not introduce bias or intentionally deceive when presenting our data, we completed our duties. However, data are not naturally occurring phenomenon. The world does not spontaneously quantify, curate, or data-mine itself. Rather, the process of observing the world and quantifying it is a political act, and deserves ethical consideration~\cite{barocas2017engaging}.

Heidegger identifies quantification as the hallmark of modern technology: the turning of things (and people) into ``standing reserves'' of resources~\cite{heidegger1954question}. A river is not just a flowing thing to be admired, it's a certain amount of megawatts of power if connected to a hydroelectric dam. An acre of forest is not just a scenic location, but a reserve of charcoal and lumber and so on. Modern technological systems are not (just) alienating, but an entire reframing of how we relate to the world around us in terms of exploiting and utilizing resources. And people are, of course, no exception. The collection of mass data about people is a way of turning them into a standing reserve (of ad revenue, of content creators, of soldiers, of bodies).

This collection of data, and the distillation of people into data, has tremendous political power. Gottfried Achenwall coined the term ``\emph{Statistik}'' to be the ``science of the state'' in his 1752 work \emph{Constitution of the Present Leading European States}. The collection of vital statistics was initially intended to be undertaken by the state for such organizational purposes as determining the size of a tax base, or the amount of trees available for naval vessels. This initial data collection was by no means apolitical: the proper data set can help start wars (Fig. \ref{fig:paine}). Nor has this centralized and  political use and meaning of statistics disappeared in the digital age: one of the first uses of computing machines to process statistical population data were the machines that IBM's subsidiary Dehomag developed for the Nazi regime, which were used to expedite and support the Final Solution~\cite{black2001ibm,dillard2003professional}. The relative emotional distance of collecting and reporting on \emph{data} (as opposed to managing and reporting on \emph{people}) arguably contributed to the uniquely \emph{bureaucratic} horrors of the Holocaust~\cite{ward2016deadly}, and to what Hannah Arendt refers to as the ``banality of evil''~\cite{arendt2006eichmann}. 

Conversely, \emph{refraining} from collecting data likewise has political and ethical consequences. Within academia, the convenience sampling of so-called WEIRD populations (Western, Educated, Industrialized, Rich, and Democratic) constrains the broader applicability of findings~\cite{henrich2010most}, and excludes populations from consideration in later designs. This imbalance in data collection can result in unequal outcomes, as with the example of the over-representation of white faces in computer vision benchmarks resulting in commercial products that fail to accurately detect or model the faces of people with darker skin~\cite{buolamwini2018}. Absence of data can be engineered for political ends: the Trump administration's attempt to add a question about citizenship to the U.S. census is likely an attempt to dissuade non-citizens from answering the census in fear of retaliation~\cite{wines2018}, and so therefore to guide the distribution of state resources away from areas with larger immigrant populations.

There is no such thing as an objective view from nowhere: rather, knowledge is \emph{situated}~\cite{haraway1988situated} within our perspectives and circumscribed by the limits of our experience. Therefore, data are not neutral and objective facts about the world--- there is no such thing as ``raw'' data~\cite{gitelman2013raw}. Data are always collected or processed by \emph{someone}, for some aim.  Often the work that goes into collecting and structuring data is made invisible~\cite{d2016feminist,datalabor}. Often, too, are the purposes for which these data are collected and used given less importance than data as an abstract puzzle to be solved or a collection of insights to be gathered. The emerging field of ``critical data science''~\cite{dalton2014does} seeks to examine how data reinforce or challenge systems of power, and to ``undo''~\cite{boyd2016undoing} assumptions that collecting more data inevitably results in an increase in efficiency or a decrease in bias.

\section{Against the Neutrality of Visualization}

\begin{figure}
\centering
\begin{subfigure}[b]{0.7\columnwidth}
	\includegraphics[width=\textwidth]{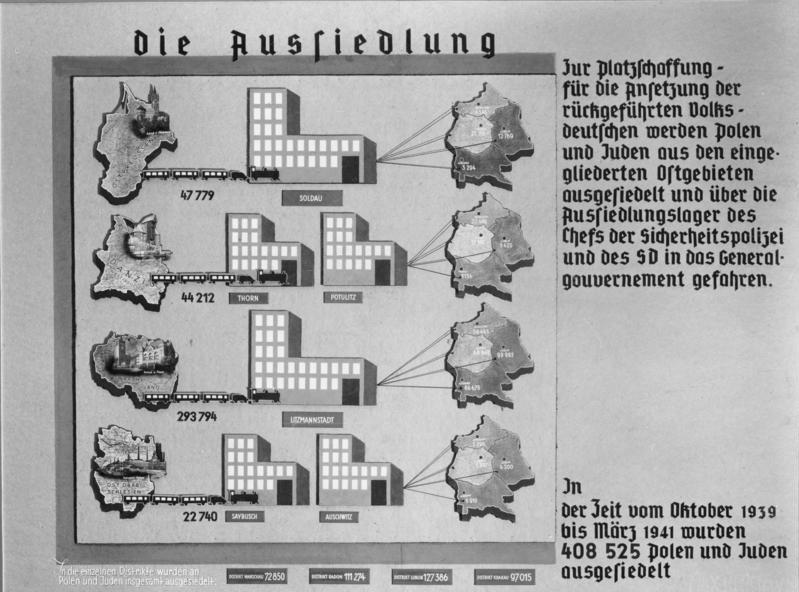}
     \caption{Source-destination map~\cite{krajewksy1940heimsinreich}}
     \label{fig:source}
\end{subfigure}
\begin{subfigure}[b]{0.7\columnwidth}
	\includegraphics[width=\textwidth]{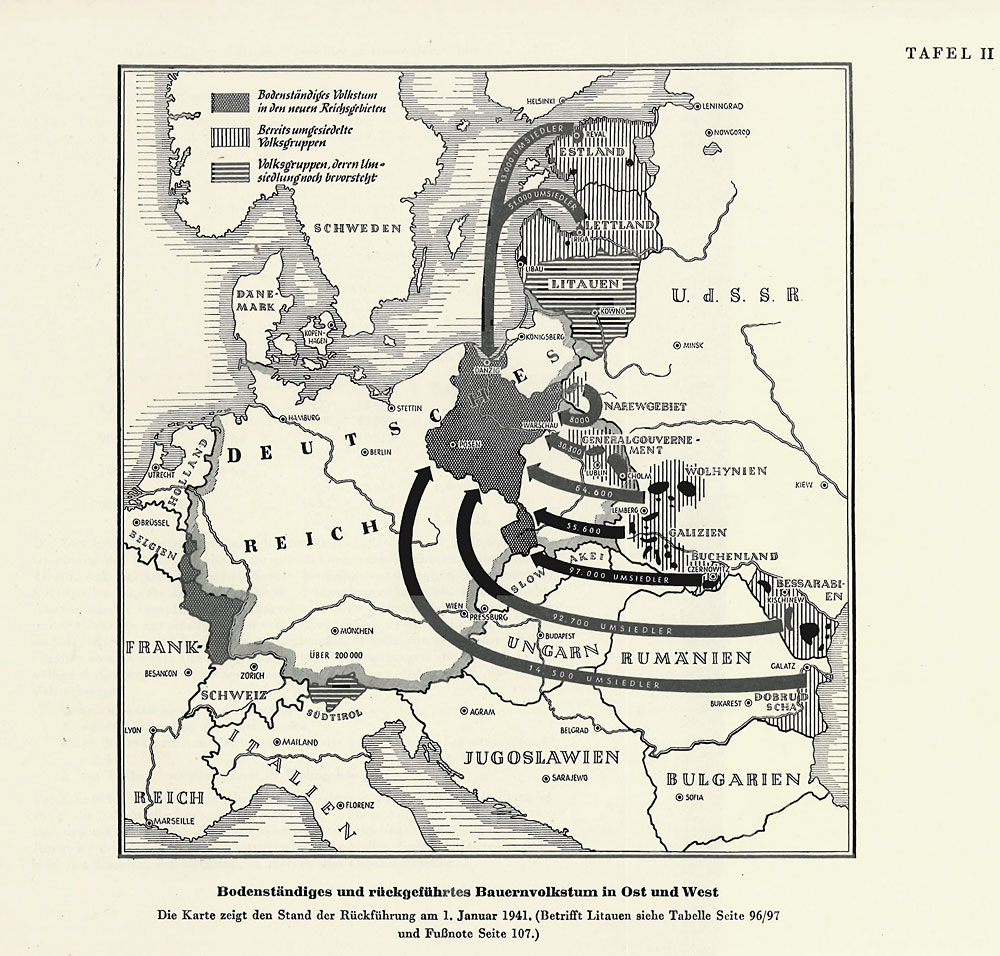}
     \caption{Flow map~\cite{meyer1942landvolk}}
     \label{fig:flow}
\end{subfigure}
\caption{Two visualizations from the Nazi regime's ``Heim ins Reich'' (Home to the Reich) campaign. This campaign was meant to promote the resettlement of ethnic Germans from other parts of Europe to newly conquered territories in Poland. The first map merely mentions that the existing Polish and Jewish population will be resettled; the second map does not mention them at all. Also invisible are the original borders of the annexed Polish state.}
\label{fig:heiminsreich}
\end{figure}

Well-designed visualizations are often conceived of as clear depictions of objective data. Drucker~\cite{drucker2012humanistic} views this framing as particularly dangerous:
\begin{quote}
While it may seem like an extreme statement, I think the ideology of almost all current information visualization is anathema to humanistic thought, antipathetic to its aims and values. The persuasive and seductive rhetorical force of visualization performs such a powerful reification of information that graphics such as Google Maps are taken to be simply a presentation of ``what is,'' as if all critical thought had been precipitously and completely jettisoned.
\end{quote}
In other words, visualizations often depict data as a \emph{given}, a collection of facts about the world that brook no argument or disagreement. Visualizations are often used as part of a rhetorical appeal to the authority and expertise of the people communicating the data~\cite{richards2003argument}, and can stifle critical or contradictory voices who do not have their own data sets to point to. Even the language we use to discuss and critique visualizations can echo implicit biases and inequalities present in society at large~\cite{hill2016visualizing}. Designers often exclude representation of factors like the uncertainty of the data or the variability of forecasts for reasons of complexity, scope, or anticipated innumeracy in the audience~\cite{boukhelifa2009uncertainty,greis2017designing}, which can contribute to the perception that the data are immutable truths about the world, rather than designed artifacts representing one flawed, incomplete, and potentially idiosyncratic set of structured observations. The clean lines and structured layouts of traditional visualizations communicate authority and certainty in implicit but measurable ways~\cite{kennedy2016work}. While visualizations can be used to promote exploration and further questioning (as in the ``martini glass''~\cite{segel2010narrative} structured narrative visualization), often designers must use unconventional designs to promote self-critique or skepticism~\cite{wood2012sketchy}.

Another concern is that data visualization, by presenting the data (rather than the people behind the data), can result in ``cruel'' and ``inhuman''~\cite{dragga2001cruel} charts. That is, by treating a chart of casualty figures as no different qualitatively than a chart of employment statistics, visualizations can hide the ethical and human suffering underlying the data. The infographics of the Nazi regime are a particularly heinous (but by no means unique) example of this erasure. For example Fig.~\ref{fig:heiminsreich} shows the colonization of conquered lands in full detail while relegating information about the forced resettlement and likely death of the original occupants to a caption. 

Visualization creates an inherent separation between the people impacted by the data and the people consuming the data. The abstraction, quantification, and digital presentation creates what Baudrillard calls ``virtualization''~\cite{baudrillard1995gulf}: an air of unreality about the needs and suffering of people of concern. Likewise, Cairo~\cite{unempathic2015} mentions that ``I am just very skeptical to the idea that data visualization is a medium that can convey (or even care about conveying) or increase `empathy','' and recent experiments by Boy et al.~\cite{boy2017showing} suggest that even designs where the human component of data are made more prominent can fail to significantly impact our empathy with human suffering.

\begin{figure}
\centering
\includegraphics[width=.6\columnwidth]{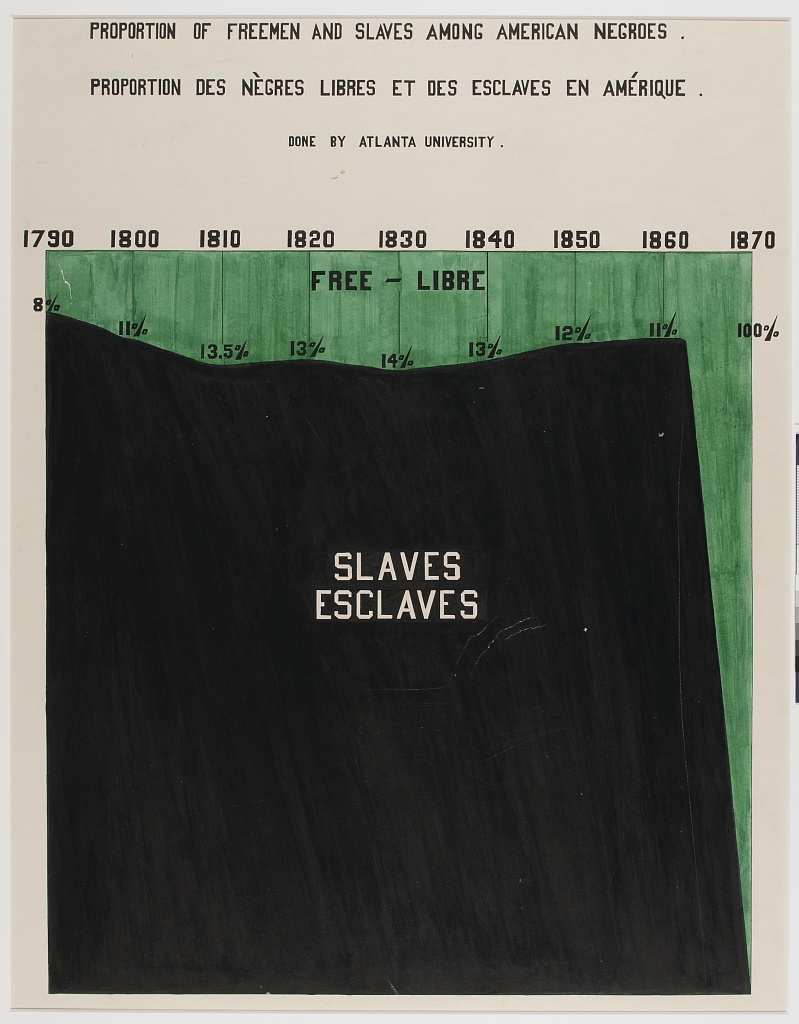}
\caption{A visualization by W.E.B. Du Bois for the 1900 Paris Exhibition\protect\cite{dubois1900}. Despite being relatively straightforward charts without much commentary, Du Bois intended that these visualizations depict the progress and dangers of the African-American population\protect\cite{lewis2010small} for a moral and political purpose.}
\label{fig:dubois}
\end{figure}


All visualizations are rhetorical, and have the power to potentially persuade~\cite{pandey2014persuasive}. Minor choices in how these charts are designed and presented can control the message that people take away~\cite{hullman2011visualization}, occasionally without conscious knowledge: e.g., the biasing title of a visualization may not be recalled, but can still measurably impact the remembered contents of a chart~\cite{kong2018frames}. 

\begin{figure*}
\centering
\begin{subfigure}[b]{0.7\columnwidth}
	\includegraphics[width=\textwidth]{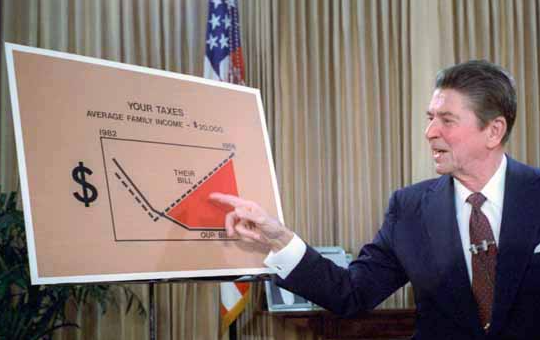}
     \caption{Ronald Reagan uses a chart as part of a public address from the Oval Office in support of the Economic Recovery Tax Act of 1981, showing the difference between the Republican and Democratic tax cut plans.}
     \label{fig:reagan}
\end{subfigure}\hspace{2em}
~
\begin{subfigure}[b]{0.7\columnwidth}
	\includegraphics[width=\textwidth]{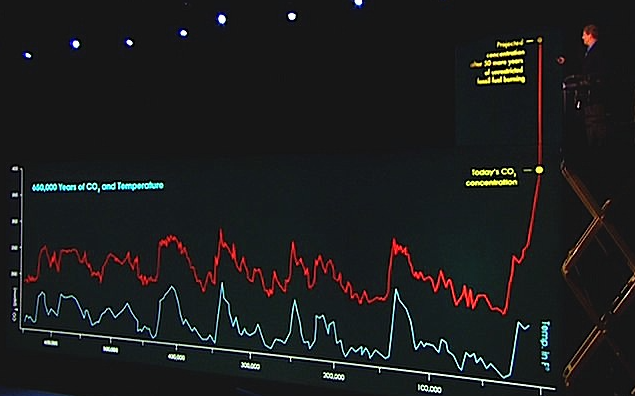}
     \caption{Al Gore uses a chart (and a scissors lift) as part of his movie \emph{An Inconvenient Truth} to show the connection between temperature and carbon dioxide, and the unprecedented scale of recent increases.}
     \label{fig:gore}
\end{subfigure}
\caption{Charts lend authority and the perception of objectivity to arguments. Designers of all charts, not just infographics or educational graphs, must be mindful of how visualizations can be used to persuade.}
\label{fig:politicalvis}
\end{figure*}

Visualization researchers may attempt to sidestep the rhetorical power of charts by separating visualizations into genres of infographics (that are meant for general audiences and can be used for persuasion) and statistical graphics (that are meant for experts and are actively discouraged from having adornments or embellishments~\cite{bateman2010useful}). However, relatively unadorned visualizations in the style of statistical graphics have a long history of use by politicians to bolster their arguments (as in Fig \ref{fig:politicalvis}).

Likewise, visualizations do not have to be explicitly placed in a political or argumentative context in order to be intended as persuasive. For instance, while the chart in Fig. \ref{fig:dubois} may appear to be a statement of demographic fact, its author, civil rights activist W.E.B. Du Bois, intended it to implicitly function as part of an argument about the status and trajectory of African-Americans ~\cite{du1926criteria}:
\begin{quote}
Thus all art is propaganda and ever must be, despite the wailing of the purists. I stand in utter shamelessness and say that whatever art I have for writing has been used always for propaganda for gaining the right of black folk to love and enjoy. I do not care a damn for any art that is not used for propaganda. But I do care when propaganda is confined to one side while the other is stripped and silent...
\end{quote}
In our own work, the assumption that attempting to persuade with visualizations is only the goal of the propagandist, and that scientific visualization and statistical graphics are therefore above such considerations, cedes rhetorical ground to the groups that do not have such scruples. 

In response to these issues, and drawing on similar concerns and methods from the ``critical cartography''~\cite{crampton2006introduction,muehlenhaus2013design} movement in GIS, D\"ork et al. have called for a ``critical infovis''~\cite{dork2013critical} movement with the goal of making explicit the values and politics of visualizations. Likewise, D'Ignazio and Klein articulate the notion of ``data feminism''~\cite{dignazio2019draft} where the power imbalances in the process of designing and deploying data visualizations are centered.

\section{Concerning Trends in Visualization Research}
There are several emerging areas of interest in the visualization community where work (or the lack of work) is a cause for concern. These areas are places where \emph{power} and \emph{responsibility} are being allocated in ways that could lead to unethical or irresponsible practice and outcomes. While a full review of all topics of visualization research, and their associated ethical considerations, is out of the scope of this paper, I selected these areas as representing ongoing areas of research where there are values and virtues in \emph{conflict}. That is, emerging topics where there may not be a single clear path forward (as in rule-based deontological ethics), but where researchers will have to balance and cultivate opposing ethical principles (as in virtue ethics~\cite{hursthouse1999virtue}). Virtue ethics does not generate prescriptive rules to follow or objective measures of success ~\cite{louden1984some}. Rather, this framing suggests mutual (occasionally conflicting) values to cultivate. 

I conclude each topic with a list of \emph{design dilemmas}: open-ended expressions of ethical implications that might arise from visualization research in these areas.

\subsection{Automated Analysis}

\begin{figure}
\centering
\includegraphics[width=.95\columnwidth]{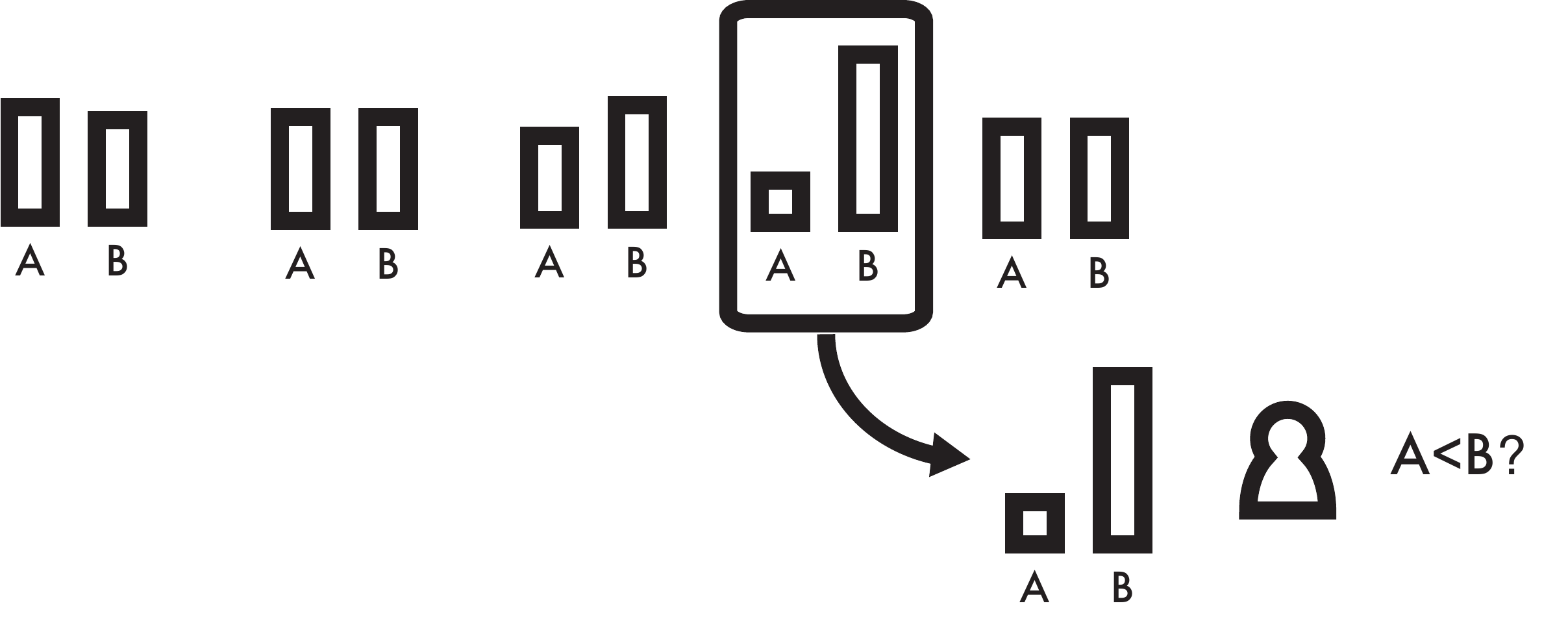}
\caption{Systems that seek to automatically locate ``insights'' in datasets can save time for users, and assist users without strong backgrounds in statistics. However, they can promote noise over signal, and lead to unjustified conclusions. How do we empower users without supporting potentially dangerous decision-making?}
\label{fig:automatic}
\end{figure}

One primary goal of visualization is the affordance of ``insights'': complex, deep, qualitative, unexpected, and relevant~\cite{north2006toward} revelations. In order to support insights, systems are beginning to explore the concept of automatic recommendations and analyses~\cite{vartak2015s,zhu2002automated}. The promise of these methods is that analysts can instantly discover important relationships in data, without having to spend many hours exploring trivial or uninteresting patterns.

However, analytics systems (or their consumers) may lack the statistical tools to \emph{validate} these insights. Therefore, analysts can frequently come away with conclusions from visualizations that are empirically false or statistically unsupported~\cite{binnig2017toward,zgraggen2018investigating}. Automatic methods can exacerbate this problem~\cite{binnig2017toward}, and create what Pu and Kay call ``p-hacking machines''~\cite{pugarden}.

Unfortunately, ``p-hacking machines'' are alluring from an end-user perspective. Finding \emph{something} is a better user experience than finding \emph{nothing}. People may lack the statistical expertise to properly make use of factors that might contextualize the importance of patterns in visualizations, like confidence intervals~\cite{belia2005researchers} or probability information~\cite{micallef2012assessing}. Very few visual analytics systems guide users not just to interesting data, but also through the process of analyzing such findings statistically~\cite{wacharamanotham2015statsplorer}. Fewer still take into account decision-making biases and attempt to correct for analytical paths not taken~\cite{wall2017warning}.

An ethical concern with this research is therefore that we are enabling bad behavior (unjustified, incorrect, and potentially damaging conclusions from data) without adequate care for the people that can be harmed by decisions based on these conclusions, or adequate understanding about the literacy and capabilities of the people who we are empowering with these automated tools.

A further concern is that the our interactive systems only exacerbate the ``garden of forking paths''\cite{gelman2013garden} problem that has contributed to the replication crisis in the sciences. Visualization research itself has many of the same issues as problematic work in other fields~\cite{kosara2018skipping}. By not creating robust ways of visualizing findings we therefore risk our own credibility as well.

However, automatic insights, by allowing people to quickly discover important facets of their data, can empower people without the time or expertise to discover these findings alone. Systems with excessive guidance or constraints also reduce the agency of the user. There is therefore a potential conflict between democratizing data analytics and promoting statistically sound decision-making (Fig. ~\ref{fig:automatic}).

\textbf{Design Dilemmas}: How much \emph{guidance} should analytics systems provide to users? How \emph{prescriptive} should such systems be in forbidding or advising against actions that are likely to lead to statistically spurious conclusions?

\subsection{Machine Learning}

\begin{figure}
\centering
\includegraphics[width=.8\columnwidth]{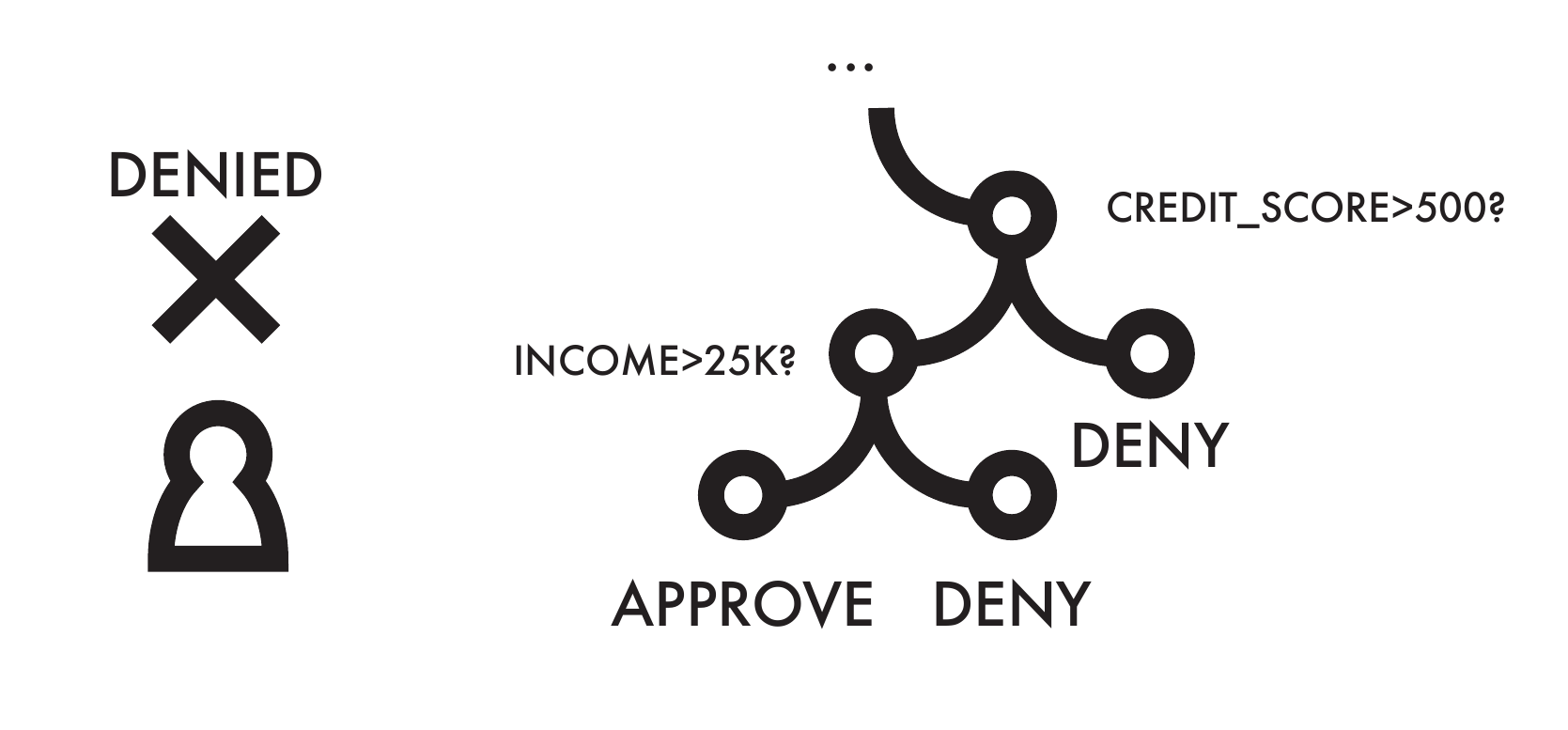}
\caption{Visualizing Machine Learning models creates a conflict between \emph{transparency} in decision-making, and managing the \emph{complexity}. This trade-off also appears to come at the expense of \emph{accuracy}. Is it more important to have a more understandable model, or a more accurate one?}
\label{fig:mlvis}
\end{figure}

Machine learning methods are powerful tools for structuring and making predictions with data, and are present in many critical areas of our society, from finance to college admissions. The resulting models are often opaque, and fail to gracefully allow appeals from people who have be wrongfully or prejudicially categorized~\cite{o2016weapons}. We have a moral duty (and, in some cases, a legal duty~\cite{wachter2017counterfactual}) to communicate decision-making based on ML to the populations that are impacted by it. Communication in this way provides much needed context for decisions that seem misguided or callous, and allow those impacted by the decisions to appeal their decisions or seek better outcomes~\cite{lecher2018}. Legal scholars such as Citron have argued for the right to ``Technological Due Process''~\cite{citron2007technological} in the face of opaque algorithmic decision-making.

Despite this obligation, and the historical positioning of visualization as a way of presenting statistical information to wider audiences, much of the prominent work on visualizing ML focuses on expert users~\cite{talbot2009ensemblematrix,wongsuphasawat2018visualizing}. An ethical concern is that we are therefore empowering the creators of ML models, but are not empowering the people affected by these models. That is, we are not focused on transparently communicating why a particular model made a choice (about one's eligibility for a loan, or eligibility for parole) to audiences without deep statistical expertise. 

Venues such as the 2018 Workshop on Visualization for AI Explainability (\url{http://visxai.io/}) are beginning to collect scholarship in this area, and online platforms such as Distill (\url{https://distill.pub/}) are beginning to collect public-facing ``explainers'' of ML concepts, but currently there are no standard methods and few success stories of visually communicating algorithmic decisions to the general audience. Even the very definition of what it means for an ML model to be ``interpretable'' is ill-defined and sometimes contradictory~\cite{lipton2016mythos}. On the ML side, work on explainability is often considered in terms of numerically representing the contribution of particular features ~\cite{lundberg2017unified,ribeiro2016should}, despite the fact that long lists of feature contributions may be difficult to interpret, or fail to speak to the domain expertise of the analyst. Methods in common use in the field, such as saliency maps~\cite{2017arXiv171100867K} or model prototypes~\cite{kim2016MMD}, are often poor conceptual models of ML behavior. Optimizing for human understanding of models often requires empirical testing and optimization independent of the modeling itself~\cite{chang2009reading}.

Simple models may be more \emph{explainable}~\cite{gleicher2013explainers}, but they are often less \emph{accurate}. The values of transparency and utility may therefore be in conflict (Fig. \ref{fig:mlvis}). We want to give the people impacted by our models the opportunity to correct errors, identify points of unfairness, and in general have agency in the decisions that affect them. On the other hand, reducing complexity to afford explainability could result in performance losses that result in worse outcomes. Likewise, there are costs even for successful explanations of ML models and decision-making. Bad actors can game the system at the expense of those who are participating fairly, as with the ``cabal''~\cite{jeong2018} of romance writers who engaged in a number of questionable activities (such as self-plagiarizing and misleading advertising) in order to consistently appear at the top of Amazon's ranking algorithms. Full transparency in models, especially in models built from demographic data, can also compromise the privacy of those who have had their data collected (as in social network data~\cite{wang2018graphprotector}).

\textbf{Design Dilemmas}: How much \emph{abstraction} or \emph{approximation} should we use when communicating complex ML models? What \emph{standards} or \emph{expectations} should we cultivate when choosing which parts of algorithmic decision-making to display?

\subsection{Provenance}

\begin{figure}
\centering
\includegraphics[width=.7\columnwidth]{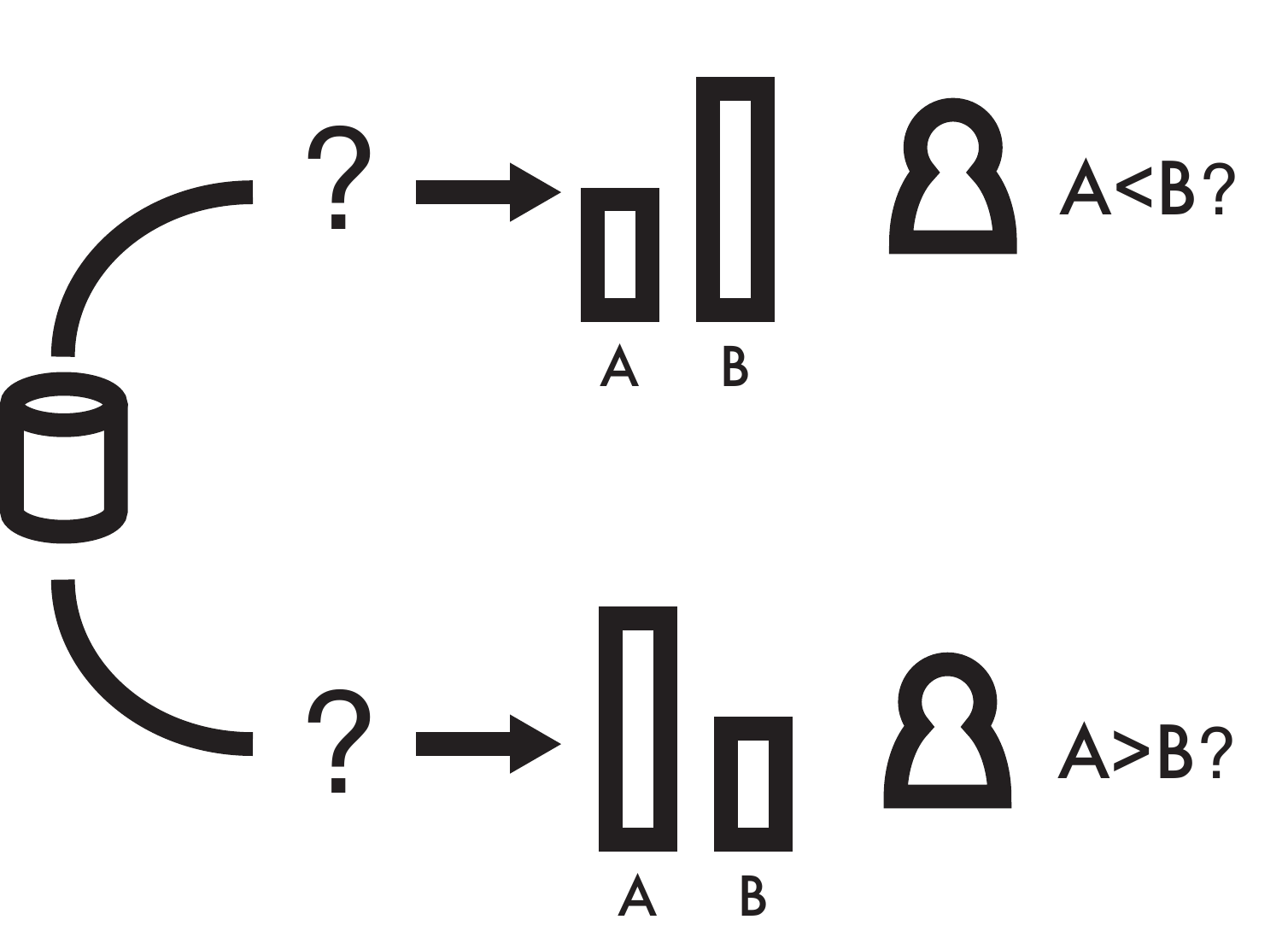}
\caption{When we visualize the \emph{end result} of a visualization design, but not the \emph{process} by which it was created, we risk propagating false, misleading, or unreproducible findings. On the other hand, showing too many extraneous details may weaken the rhetorical impact, and increase the complexity, of visualizations. How do we use data to convince people, but without taking away agency? }
\label{fig:provenance}
\end{figure}

Visual analytics systems are increasing in both complexity and importance. Combined with the ``garden of forking paths'' problem mentioned above, this large number of potential actions means that it is becoming increasingly difficult to articulate exactly what steps an analyst took in order to produce a particular chart or to arrive at a particular conclusion. This need becomes even more important as analytical systems become more tightly integrated with machine learning, which can be non-deterministic in its output, or highly dependent on hyperparameters in its input.

An unmet ethical challenge in visualization is therefore to visualize the provenance of data and decision-making. Communicating the decisions that an analyst took, and affording different decisions, is a key component of both affording criticism and supporting transparency in data-driven decision-making. Much of visualization work is instead focused on affording exploration and analysis, rather than communication of how this exploration and analysis was performed~\cite{kosara2013storytelling}. There is initial work in increasing the transparency of visualizations: systems like Vistrails~\cite{callahan2006vistrails} and Hindsight~\cite{feng2017hindsight} represent initial steps at visualizing scientific workflows and user viewing histories, respectively. Similarly, ``literate visualization''~\cite{wood2018design} has the goal of making the design decisions that lead to a final visualization documented and transparent. However, very few visual analytics systems are built with the goal of analytical transparency in mind.

Notebook-style interfaces such as Jupyter and Observable and other literate programming environments such as R markdown represent important ecosystems for the transparent communication of analyses, but require coding or scripting expertise to construct. In contrast, popular visual analytics systems (such as Tableau, PowerBI, and Spotfire) heavily rely on GUIs and do not require coding expertise to use. There is therefore a gap between ease of analysis and ease of documentation, further muddying the waters between exploratory and confirmatory analytics.

A connected challenge is the rhetorical one of how to convince viewers not to be taken in by unreliable data or information (such as ``fake news''~\cite{allcott2017social}) or how to mitigate cognitive biases in decision-making~\cite{dimara2018}. Here there is an ethical balance between supporting the agency and desires of the viewer (who may not appreciate being intentionally guided away from the information they want to see) and the desire to communicate information that is both correct and useful. Visualizing provenance information and making explicit the analytical choices made by a system is one way of navigating between these two competing values (Fig \ref{fig:provenance}).

\textbf{Design Dilemmas}: How, and how many, \emph{alternate design or analytical decisions} should we surface to the user? Should we \emph{audit} or \emph{structure} the provenance of a visualization in order to surface irregularities?

\section{What Are Our Obligations As Visualization Researchers?}

Visualization, especially visualization research, operates at the intersection of science, communication, and engineering. We have certain ethical obligations as scientists (for instance, to avoid breaches of consent and excesses of harm laid out in codes of research conduct such as the Declaration of Helsinki~\cite{doi:10.1001/jama.2013.281053}. Likewise, we have ethical obligations as engineers (for instance, to avoid doing shoddy work or work dangerous to the public good) as laid out in professional codes of conduct such as the ACM's code of ethics~\cite{anderson1992acm}. Lastly, insofar as we are presenting data to the public, we have ethical obligations as journalists (for instance, to issue corrections and disclose conflicts of interest), as laid out in codes of conduct such as the SPJ's code of ethics~\cite{farley2014spj}. 

Beyond the obligation of the component parts of visualization praxis, we also have obligations in that we have a great deal of power over how people ultimately make use of data, both in the patterns they see and the conclusions they draw~\cite{correll2017black}. We are often the first and only contact a person might have with an underlying store of data. This gives us special access to impacted populations, and special responsibilities as we control the curation, presentation, rhetorical content of the visualizations we create. Visualizations systems we create also embody design principles concerning democratization, transparency, clarity, and automation that lend their use a unique moral signature.

In the following subsections I will present three ethical challenges of visualization work, related to visibility, privacy, and power. I will briefly describe how visualization work impacts these realms, and suggest some virtues or related principles that can ameliorate the negative impact of visualization work in these spheres. In the spirit of virtue ethics, I do not view these principles as unimpeachable or absolute. Therefore, I end each section with a list of potential caveats, where adherence to these principles can have unwanted ethical impacts, or where there exist virtues whose cultivation may directly conflict with the principles I propose. 

\subsection{Make the Invisible Visible}
There are many potentially invisible aspects of a visualization (Fig. \ref{fig:teaser}). These non-visualized components such as the choices (and labor) that went into collecting and curating the data, or the populations that could be impacted by the decisions made by viewers of these visualizations, have a non-trivial impact on the good (or harm) that a visualization can do. I echo the view of D\"ork et al.~\cite{dork2013critical} that we have a responsibility to make the invisible visible:

We ought to \textbf{visualize hidden labor}. Properly acknowledging and rewarding people for their labor is a key component of fairness. Certain kinds of labor (especially those performed by marginalized groups) are under-represented or under-valued in our current schemes of commodification or valuation. For instance, the ``emotional labor''~\cite{hochshild1983managed} of people in service and nursing-related professions is often overlooked. Echoing D'Ignazio and Klein~\cite{d2016feminist}, I believe that the labor that goes into collecting, curating, and archiving data, and the further work of analyzing the data, is often invisible in visualizations. Visualizations are often presented as finished products, with the steps in their construction (and alternative steps not taken) hidden from view. Beyond fairness in attribution, making this labor visible is also of benefit to the progress of the field as a whole. Making the labor of analysis and data prep visible will contribute to reproducibility and openness, and will also facilitate in the creation of established standards. Making the work of design and user research visible (including surfacing intermediate or failed prototypes) will serve as points of inspiration for other designers, or warnings about potential unfruitful avenues of effort. 

We ought to \textbf{visualize hidden uncertainty}. Uncertainty is an inescapable component of data collection and analysis, yet it is often hidden from the end user for reasons of complexity or anticipated literacy. Recent work in novel visualizations of uncertainty has specifically targeted general audiences~\cite{greis2017designing,kay2016ish} and shown that the ability of the general public to make use of uncertainty information can be quite high given the right kind of visual presentation. Weather data, personal informatics, and electoral polling are all examples where the general public is presented with data with an inescapable component of uncertainty. The way that this uncertainty is presented can have measurable impacts on decision-making. For instance the perceived risk of hurricanes is impacted by how their paths are depicted~\cite{ruginski2016non}, and voter turnout in elections can be impacted by the race's perceived closeness~\cite{ashworth2006everyone}, which in turn can be impacted by how the polling data are presented~\cite{correll2014error}. To encourage better decision-making (or more uncertainty-aware decision-making), we must investigate the design space of uncertainty visualization and how to measure its impact on our audiences, just as we would also wish to measure task speed or accuracy~\cite{hullman2018pursuit}.

We ought to \textbf{visualize hidden impacts}. The ACM Future of Computing Academy has called for writers of academic papers to make the potential impacts (especially the potential negative impacts) of academic work to be made explicit~\cite{fca2018}. Visualization work often focuses on the positive aspects of a system (for instance, its ease of use, or the speed or accuracy with which analysts conduct their tasks), but rarely on the potential of these systems for harm or misuse. For instance, what harm could be done using the new classes of insights afforded by the visualization system? What groups or individuals may be impacted by the work but do not fit the data model in use? Are there immoral or predatory domains that could make use of the proposed techniques? Beyond these potentially extreme examples of negative impacts, there are potential negative impacts even for visualizations that have been judged to have succeeded in their goals. For instance, will people be harmed if and when the system ceases to be supported (a common occurrence for systems developed in academia)? What are the opportunity costs (in terms of training, collaboration time, and student or developer labor) associated with the system? Do all of the components of the ultimate value of the visualization~\cite{van2005value} have a positive sign, from a consequentialist perspective?

\textbf{Caveats}:
Visualizations are already complex and multifaceted artifacts, and designers must frequently struggle with the comprehensibility of their designs and the literacy of their audience. Visualizing additional facets of the data, even for laudatory reasons of transparency and accountability, only exacerbate these issues. In addition to making visualizations more difficult to interpret (and so limiting their audience), depicting uncertainty and counter-narratives without proper care can weaken the rhetorical impact of visualizations. \textbf{Managing complexity} is therefore a virtue in design that can be in direct opposition with the desire to visualize the invisible.


\subsection{Collect Data With Empathy}
Ceglowski places many of the ills of the current internet at the hands of ``investor storytime''\cite{ceglowski2014internet}: the fable that existing data models are not quite up to the task of performing feats like microtargeting ads or predicting user behavior, but that they will be \emph{if only we collect more, and more personal} data. This goal results in an increasing pressure to collect as much data as possible to improve models or build bigger pictures or discover more context. The resulting ecosystem of omnipresent data collection means that it is becoming easier and easier to breach the privacy of users with relatively little effort~\cite{vines2017exploring}. Visualization is well-positioned to explicitly push against the pressure to collect more and more data by better communicating data's value and impact to analysts. We also have the option of contextualizing and curating data in a way that is respectful of our larger sets of shared values:

We ought to \textbf{encourage ``small data.''} boyd and Crawford claim that ``bigger data are not always better data''~\cite{boyd2012critical}. The collection of additional data is not just an expenditure of time and resources, it can also intrusively erode the privacy and agency of the people who are subject to this data collection. Even if a particular dataset is used for a population's benefit, merely setting or eroding expectations of how much data one ``needs'' to perform analysis can have negative repercussions for vulnerable populations in the future. Designers of visualizations and analytics systems should be able to communicate how much data is ``enough,'' and condition analysts to accept tradeoffs of accuracy or certainty in exchange for concision and protection.

We ought to \textbf{anthropomorphize data}. Much of visualization's power comes from the power of abstraction, but this creates a gap between populations and how they are represented in visualizations: quantization and virtualization of human beings stymies empathy. Designers must attempt to cross this gap between map and territory, especially for visualizations with high moral stakes such as those concerning human suffering. Proposed solutions to recenter human beings in data visualizations using person-shaped glyphs has not been shown to produce any additional empathic response~\cite{boy2017showing}, but including actual human beings in visualizations can help communicate complex phenomena~\cite{slobin2014} and contribute to human interest in, and memorability of, visualizations~\cite{borkin2013makes,borkin2016beyond}. Visualizations designers may have to borrow techniques from journalism and rhetoric, and propose novel designs or interventions, in order to foster empathy and spur action using visualizations. 

We ought to \textbf{obfuscate data to protect privacy}. Visualization designers often have privileged access to sensitive datasets, and are then charged with communicating these datasets to wider public spheres. The privacy and consent of the people whose data we collect or measure is therefore of paramount ethical importance. People are often unfamiliar with how much data is really collected or collectible through public APIs or other services. These inadvertent exposures of data, combined with the ability of visualizations to highlight previously unseen patterns and trends, can result in severe breaches of confidentiality and privacy, as with the recent use of fitness tracking company Strava's user heatmap and public API to identify the locations and internal layouts of U.S. military bases~\cite{hsu2018}. Preserving the privacy of the people in our dataset may involve novel designs~\cite{dasgupta2011adaptive} or interactive workflows~\cite{wang2018graphprotector}. Both of these classes of techniques involve aggregating, fuzzing, or otherwise restructuring data to preserve privacy. A related component of this obfuscation is then communicating the upper limit of accuracy or detail to analysts.

\textbf{Caveats}:
Restricting the type and amount of data that we collect has a direct impact on the quality and scope of our analyses. It may be laudatory to avoid collecting unnecessary information, but these seemingly irrelevant fields can result in serendipitous discoveries that might not have otherwise been possible. Limiting the scope of data collection also entails a form of selection bias that could result in biased or unjust conclusions arising from a lack of context. Aggregation to preserve privacy also can result in seemingly contradictory conclusions such as Simpson's Paradox. Our obligation to \textbf{provide context and analytical power} can therefore stand in direct opposition to the empathic collection of data.

Likewise, our empathic judgments can be biased or otherwise fraught. For instance, Nagel's concept of \emph{moral luck}~\cite{williams1976moral} notes the role that blind chance plays on our moral judgments. For instance, attempted murderers receive lighter sentences than murderers who succeed in killing their victims, even if their intent and actions were identical. Similarly, Bloom~\cite{bloom2017against} argues that our visceral empathic reactions can be misused in service of violent or discriminatory ends. Cultivating our \emph{sympathy} may therefore cause us to come in conflict with \textbf{institutional fairness}. Patil et al~\cite{patil2018ethics} argue for a checklist-based system of ethical reflection to circumvent these complexities.

\subsection{Challenge Structures of Power}
Satirist Peter Finley Dunne suggested that one of the jobs of newspapers is to ``comfort the afflicted'' and ``afflict the comfortable.'' Data visualizations do not often achieve these goals. The process of collecting data still requires money, time, access, and storage, which inherently gives the advantage and priority to governments and corporations with access to those resources. Many of the resources for performing academic research likewise originate from powerful governments or corporations. The academic visualization focus on esoteric or otherwise complex datasets means that the intended audience of a visualization is often those with high scientific, mathematic, and visual literacies rather than wider and more general audiences. Visualization work should be concerned with imbalances in power, and focus on distributing power in more equitable ways, and to more ethical ends:


We ought to \textbf{support data ``due process''}. Citron's~\cite{citron2007technological} notion of a technological due process highlights the danger that automated and data-driven decision-making has on our norms of decision-making and procedure. Datasets only imperfectly capture important information relevant to decsion-making, but may then become reified by algorithms or visualizations into unappealable assertions about the state of the world. People impacted by these systems have the same rights as they do with other forms of decision-making, and deserve some say in how they are (or are not) represented. Legal frameworks such as the EU's GDPR are beginning to codify rights such as ``the right to explanation,'' ``the right to privacy,'' and the ``right to be forgotten.'' However, these existing laws capture our intuitive notions about these rights only imperfectly~\cite{wachter2017counterfactual}. Our ethical obligations therefore may or may not reflect the letter of the law. Compounding this issue is that many of these algorithmic systems are of sufficient size or complexity that there are no clear procedures for visualizing them, especially for audiences that lack statistical expertise or extensive context about the domain. We therefore also have design research and pedagogical responsibilities to ensure that we are giving people agency and representation in ways that are useful and understandable.

We ought to \textbf{act as data advocates}. Visualizations have rhetorical strength and political power. Government agencies and corporations have explicit resources dedicated to the design and publication of annual reports and data reporting. Marginalized groups do not often have access to the same set of resources, and so are under-represented in data-based conversations. Conversely, groups may have financial or political interest in muddying the waters around debates of fact such as climate change or humanitarian crises. Just as it is considered laudatory to donate time or money to charitable causes, we should also donate a portion of our expertise in the presentation of information to advocate or amplify causes we believe in. This could be a relatively low-cost endeavor. For instance, many visualization papers use a similar set of standard datasets to illustrate or evaluate their designs. These datasets often have limited relevance or importance~\cite{kosara2018}. Alternative datasets about issues of current concern would also suffice to show that a system operates correctly, but could increase the visibility of ongoing injustices.

We ought to \textbf{pressure or slow unethical analytical behavior}. In response to abuses by U.S. Immigration and Customs Enforcement (ICE) agency, Amazon employees circulated an internal memo asking CEO Jeff Bezos to cut ties with the agency~\cite{shaban2018}. Google researcher Jack Poulson, \emph{inter alia}, resigned from Google over ethical concerns about the design of a search engine that censors internet content in mainland China~\cite{gallagher2018,poulson2018}. Public resignations and dissent can surface perceived ethical lapses in companies, but many may lack the financial or political security to engage in such tactics. Likewise, organizational power may seek to circumvent or exclude those with ethical concerns from decision-making: Google's senior management reportedly kept its censorship project secret from the internal teams that typically engage with the ethical implications of Google's work~\cite{gallagher2018b}. Even for those outside of such organizations, the very companies or governments engaged in ethical lapses are also frequently the source of funds and exposure for visualization research both in academia and industrial research. In such cases where one is unwilling or unable to risk retaliation to stop or speak out against unethical work, other options are the intentional sabotage or slowdown of labor. Within the context of visualization work, this could be overestimation of budgets of money and time, underestimation of result quality, or unnecessary delays (say, for additional user testing or ablation studies). Such sabotage may involve a conflict between \emph{ethical}, \emph{professional} and perhaps even \emph{legal} duties, and should not be undertaken lightly.

\textbf{Caveats}:
Conspiracy theorists, political extremists, and corporate interests (such as the tobacco and oil industries) make use of the margins of discourse, counter-narratives, and doubt to advance agendas that rely on the general public discarding the opinion of experts. These bad-faith epistemologies combined with the increasingly fractal nature of academic study has resulted in what Nichols calls ``the death of expertise''~\cite{nichols2017death}: a ongoing and increasing hostility to scientific and technocratic sources of knowledge. While people and organizations collecting data may be in positions of power compared to the general public, they might be at a further power disadvantage compared to these organizations that do not want the public to have access to or comprehension of particular information. The goal of \textbf{promoting truth and suppressing falsehood} may require amplifying existing structures of expertise and power, and suppressing conflicts for the sake of rhetorical impact.


\section{Conclusion}
This work presents some of the pressing ethical considerations of visualization work, but it functions as neither a complete survey of this space nor an exhaustive and prescriptive decision criteria to guarantee that a visualization was designed or deployed ethically (given the disagreements that rational people can have over what constitutes a moral course of action, no such criterion is likely to exist). Future work requires both developing a new pedagogy for instilling the right values in visualization designers and researchers as well as post-hoc studies of the ethical impact of visualization work in the existing moral landscapes.

It is my intention that this work functions as a both a synthesis of existing critical and ethical views of data and visualization as well as a call to action to be mindful of the ethical implications of our work, and to cultivate the right values and virtues in our work moving forward.


\bibliographystyle{ACM-Reference-Format}
\bibliography{bib}


\begin{thebibliography}{108}


\ifx \showCODEN    \undefined \def \showCODEN     #1{\unskip}     \fi
\ifx \showDOI      \undefined \def \showDOI       #1{#1}\fi
\ifx \showISBNx    \undefined \def \showISBNx     #1{\unskip}     \fi
\ifx \showISBNxiii \undefined \def \showISBNxiii  #1{\unskip}     \fi
\ifx \showISSN     \undefined \def \showISSN      #1{\unskip}     \fi
\ifx \showLCCN     \undefined \def \showLCCN      #1{\unskip}     \fi
\ifx \shownote     \undefined \def \shownote      #1{#1}          \fi
\ifx \showarticletitle \undefined \def \showarticletitle #1{#1}   \fi
\ifx \showURL      \undefined \def \showURL       {\relax}        \fi
\providecommand\bibfield[2]{#2}
\providecommand\bibinfo[2]{#2}
\providecommand\natexlab[1]{#1}
\providecommand\showeprint[2][]{arXiv:#2}

\bibitem[\protect\citeauthoryear{Allcott and Gentzkow}{Allcott and
  Gentzkow}{2017}]%
        {allcott2017social}
\bibfield{author}{\bibinfo{person}{Hunt Allcott} {and} \bibinfo{person}{Matthew
  Gentzkow}.} \bibinfo{year}{2017}\natexlab{}.
\newblock \showarticletitle{Social media and fake news in the 2016 election}.
\newblock \bibinfo{journal}{\emph{Journal of Economic Perspectives}}
  \bibinfo{volume}{31}, \bibinfo{number}{2} (\bibinfo{year}{2017}),
  \bibinfo{pages}{211--36}.
\newblock


\bibitem[\protect\citeauthoryear{Anderson}{Anderson}{1992}]%
        {anderson1992acm}
\bibfield{author}{\bibinfo{person}{Ronald~E Anderson}.}
  \bibinfo{year}{1992}\natexlab{}.
\newblock \showarticletitle{ACM code of ethics and professional conduct}.
\newblock \bibinfo{journal}{\emph{Commun. ACM}} \bibinfo{volume}{35},
  \bibinfo{number}{5} (\bibinfo{year}{1992}), \bibinfo{pages}{94--99}.
\newblock


\bibitem[\protect\citeauthoryear{Arendt}{Arendt}{2006}]%
        {arendt2006eichmann}
\bibfield{author}{\bibinfo{person}{Hannah Arendt}.}
  \bibinfo{year}{2006}\natexlab{}.
\newblock \bibinfo{booktitle}{\emph{Eichmann in jerusalem}}.
\newblock \bibinfo{publisher}{Penguin}.
\newblock


\bibitem[\protect\citeauthoryear{Ashworth, Geys, and Heyndels}{Ashworth
  et~al\mbox{.}}{2006}]%
        {ashworth2006everyone}
\bibfield{author}{\bibinfo{person}{John Ashworth}, \bibinfo{person}{Benny
  Geys}, {and} \bibinfo{person}{Bruno Heyndels}.}
  \bibinfo{year}{2006}\natexlab{}.
\newblock \showarticletitle{Everyone likes a winner: An empirical test of the
  effect of electoral closeness on turnout in a context of expressive voting}.
\newblock \bibinfo{journal}{\emph{Public Choice}} \bibinfo{volume}{128},
  \bibinfo{number}{3-4} (\bibinfo{year}{2006}), \bibinfo{pages}{383--405}.
\newblock


\bibitem[\protect\citeauthoryear{Association}{Association}{2013}]%
        {doi:10.1001/jama.2013.281053}
\bibfield{author}{\bibinfo{person}{World~Medical Association}.}
  \bibinfo{year}{2013}\natexlab{}.
\newblock \showarticletitle{World medical association declaration of helsinki:
  Ethical principles for medical research involving human subjects}.
\newblock \bibinfo{journal}{\emph{JAMA}} \bibinfo{volume}{310},
  \bibinfo{number}{20} (\bibinfo{year}{2013}), \bibinfo{pages}{2191--2194}.
\newblock
\urldef\tempurl%
\url{https://doi.org/10.1001/jama.2013.281053}
\showDOI{\tempurl}
\showeprint{/data/journals/jama/929397/jsc130006.pdf}


\bibitem[\protect\citeauthoryear{Barocas and boyd}{Barocas and boyd}{2017}]%
        {barocas2017engaging}
\bibfield{author}{\bibinfo{person}{Solon Barocas} {and} \bibinfo{person}{danah
  boyd}.} \bibinfo{year}{2017}\natexlab{}.
\newblock \showarticletitle{Engaging the ethics of data science in practice}.
\newblock \bibinfo{journal}{\emph{Commun. ACM}} \bibinfo{volume}{60},
  \bibinfo{number}{11} (\bibinfo{year}{2017}), \bibinfo{pages}{23--25}.
\newblock


\bibitem[\protect\citeauthoryear{Bateman, Mandryk, Gutwin, Genest, McDine, and
  Brooks}{Bateman et~al\mbox{.}}{2010}]%
        {bateman2010useful}
\bibfield{author}{\bibinfo{person}{Scott Bateman}, \bibinfo{person}{Regan~L
  Mandryk}, \bibinfo{person}{Carl Gutwin}, \bibinfo{person}{Aaron Genest},
  \bibinfo{person}{David McDine}, {and} \bibinfo{person}{Christopher Brooks}.}
  \bibinfo{year}{2010}\natexlab{}.
\newblock \showarticletitle{Useful junk?: the effects of visual embellishment
  on comprehension and memorability of charts}. In
  \bibinfo{booktitle}{\emph{Proceedings of the SIGCHI Conference on Human
  Factors in Computing Systems}}. ACM, \bibinfo{pages}{2573--2582}.
\newblock


\bibitem[\protect\citeauthoryear{Baudrillard}{Baudrillard}{1995}]%
        {baudrillard1995gulf}
\bibfield{author}{\bibinfo{person}{Jean Baudrillard}.}
  \bibinfo{year}{1995}\natexlab{}.
\newblock \bibinfo{booktitle}{\emph{The Gulf War did not take place}}.
\newblock \bibinfo{publisher}{Indiana University Press}.
\newblock


\bibitem[\protect\citeauthoryear{Belia, Fidler, Williams, and Cumming}{Belia
  et~al\mbox{.}}{2005}]%
        {belia2005researchers}
\bibfield{author}{\bibinfo{person}{Sarah Belia}, \bibinfo{person}{Fiona
  Fidler}, \bibinfo{person}{Jennifer Williams}, {and} \bibinfo{person}{Geoff
  Cumming}.} \bibinfo{year}{2005}\natexlab{}.
\newblock \showarticletitle{Researchers misunderstand confidence intervals and
  standard error bars.}
\newblock \bibinfo{journal}{\emph{Psychological methods}} \bibinfo{volume}{10},
  \bibinfo{number}{4} (\bibinfo{year}{2005}), \bibinfo{pages}{389}.
\newblock


\bibitem[\protect\citeauthoryear{Binnig, De~Stefani, Kraska, Upfal, Zgraggen,
  and Zhao}{Binnig et~al\mbox{.}}{2017}]%
        {binnig2017toward}
\bibfield{author}{\bibinfo{person}{Carsten Binnig}, \bibinfo{person}{Lorenzo
  De~Stefani}, \bibinfo{person}{Tim Kraska}, \bibinfo{person}{Eli Upfal},
  \bibinfo{person}{Emanuel Zgraggen}, {and} \bibinfo{person}{Zheguang Zhao}.}
  \bibinfo{year}{2017}\natexlab{}.
\newblock \showarticletitle{Toward Sustainable Insights, or Why Polygamy is Bad
  for You.}. In \bibinfo{booktitle}{\emph{Proceedings of the Conference on
  Innovative Data Systems Research (CIDR)}}.
\newblock


\bibitem[\protect\citeauthoryear{Black}{Black}{2001}]%
        {black2001ibm}
\bibfield{author}{\bibinfo{person}{Edwin Black}.}
  \bibinfo{year}{2001}\natexlab{}.
\newblock \bibinfo{booktitle}{\emph{IBM and the Holocaust: The strategic
  alliance between Nazi Germany and America's most powerful corporation}}.
\newblock \bibinfo{publisher}{Random House Inc.}
\newblock


\bibitem[\protect\citeauthoryear{Bloom}{Bloom}{2017}]%
        {bloom2017against}
\bibfield{author}{\bibinfo{person}{Paul Bloom}.}
  \bibinfo{year}{2017}\natexlab{}.
\newblock \bibinfo{booktitle}{\emph{Against empathy: The case for rational
  compassion}}.
\newblock \bibinfo{publisher}{Random House}.
\newblock


\bibitem[\protect\citeauthoryear{Borkin, Bylinskii, Kim, Bainbridge, Yeh,
  Borkin, Pfister, and Oliva}{Borkin et~al\mbox{.}}{2016}]%
        {borkin2016beyond}
\bibfield{author}{\bibinfo{person}{Michelle~A Borkin}, \bibinfo{person}{Zoya
  Bylinskii}, \bibinfo{person}{Nam~Wook Kim}, \bibinfo{person}{Constance~May
  Bainbridge}, \bibinfo{person}{Chelsea~S Yeh}, \bibinfo{person}{Daniel
  Borkin}, \bibinfo{person}{Hanspeter Pfister}, {and} \bibinfo{person}{Aude
  Oliva}.} \bibinfo{year}{2016}\natexlab{}.
\newblock \showarticletitle{Beyond memorability: Visualization recognition and
  recall}.
\newblock \bibinfo{journal}{\emph{IEEE transactions on visualization and
  computer graphics}} \bibinfo{volume}{22}, \bibinfo{number}{1}
  (\bibinfo{year}{2016}), \bibinfo{pages}{519--528}.
\newblock


\bibitem[\protect\citeauthoryear{Borkin, Vo, Bylinskii, Isola, Sunkavalli,
  Oliva, and Pfister}{Borkin et~al\mbox{.}}{2013}]%
        {borkin2013makes}
\bibfield{author}{\bibinfo{person}{Michelle~A Borkin},
  \bibinfo{person}{Azalea~A Vo}, \bibinfo{person}{Zoya Bylinskii},
  \bibinfo{person}{Phillip Isola}, \bibinfo{person}{Shashank Sunkavalli},
  \bibinfo{person}{Aude Oliva}, {and} \bibinfo{person}{Hanspeter Pfister}.}
  \bibinfo{year}{2013}\natexlab{}.
\newblock \showarticletitle{What makes a visualization memorable?}
\newblock \bibinfo{journal}{\emph{IEEE Transactions on Visualization and
  Computer Graphics}} \bibinfo{volume}{19}, \bibinfo{number}{12}
  (\bibinfo{year}{2013}), \bibinfo{pages}{2306--2315}.
\newblock


\bibitem[\protect\citeauthoryear{Boukhelifa and Duke}{Boukhelifa and
  Duke}{2009}]%
        {boukhelifa2009uncertainty}
\bibfield{author}{\bibinfo{person}{Nadia Boukhelifa} {and}
  \bibinfo{person}{David~John Duke}.} \bibinfo{year}{2009}\natexlab{}.
\newblock \showarticletitle{Uncertainty visualization: why might it fail?}. In
  \bibinfo{booktitle}{\emph{CHI'09 Extended Abstracts on Human Factors in
  Computing Systems}}. ACM, \bibinfo{pages}{4051--4056}.
\newblock


\bibitem[\protect\citeauthoryear{Boy, Pandey, Emerson, Satterthwaite, Nov, and
  Bertini}{Boy et~al\mbox{.}}{2017}]%
        {boy2017showing}
\bibfield{author}{\bibinfo{person}{Jeremy Boy}, \bibinfo{person}{Anshul~Vikram
  Pandey}, \bibinfo{person}{John Emerson}, \bibinfo{person}{Margaret
  Satterthwaite}, \bibinfo{person}{Oded Nov}, {and} \bibinfo{person}{Enrico
  Bertini}.} \bibinfo{year}{2017}\natexlab{}.
\newblock \showarticletitle{Showing People Behind Data: Does Anthropomorphizing
  Visualizations Elicit More Empathy for Human Rights Data?}. In
  \bibinfo{booktitle}{\emph{Proceedings of the 2017 CHI Conference on Human
  Factors in Computing Systems}}. ACM, \bibinfo{pages}{5462--5474}.
\newblock


\bibitem[\protect\citeauthoryear{boyd}{boyd}{2016}]%
        {boyd2016undoing}
\bibfield{author}{\bibinfo{person}{danah boyd}.}
  \bibinfo{year}{2016}\natexlab{}.
\newblock \showarticletitle{Undoing the neutrality of big data}.
\newblock \bibinfo{journal}{\emph{Florida Law Review}}  \bibinfo{volume}{67}
  (\bibinfo{year}{2016}), \bibinfo{pages}{226--232}.
\newblock


\bibitem[\protect\citeauthoryear{boyd and Crawford}{boyd and Crawford}{2012}]%
        {boyd2012critical}
\bibfield{author}{\bibinfo{person}{danah boyd} {and} \bibinfo{person}{Kate
  Crawford}.} \bibinfo{year}{2012}\natexlab{}.
\newblock \showarticletitle{Critical questions for big data: Provocations for a
  cultural, technological, and scholarly phenomenon}.
\newblock \bibinfo{journal}{\emph{Information, communication \& society}}
  \bibinfo{volume}{15}, \bibinfo{number}{5} (\bibinfo{year}{2012}),
  \bibinfo{pages}{662--679}.
\newblock


\bibitem[\protect\citeauthoryear{Bundesarchiv}{Bundesarchiv}{1940}]%
        {krajewksy1940heimsinreich}
\bibfield{author}{\bibinfo{person}{Bundesarchiv}.}
  \bibinfo{year}{1940}\natexlab{}.
\newblock \bibinfo{title}{Planung und Aufbau im Osten}.
\newblock
  \bibinfo{howpublished}{\url{https://en.wikipedia.org/wiki/Heim_ins_Reich\#/media/File:Bundesarchiv_R_49_Bild-0025,_Ausstellung_\%22Planung_und_Aufbau_im_Osten\%22\,_Schautafel.jpg}}.
\newblock
\newblock
\shownote{Photograph by M. Krajewsky.}


\bibitem[\protect\citeauthoryear{Buolamwini}{Buolamwini}{2018}]%
        {buolamwini2018}
\bibfield{author}{\bibinfo{person}{Joy Buolamwini}.}
  \bibinfo{year}{2018}\natexlab{}.
\newblock \showarticletitle{When the Robot Doesn't See Dark Skin}.
\newblock \bibinfo{journal}{\emph{New York Times}} (\bibinfo{date}{Jun}
  \bibinfo{year}{2018}).
\newblock
\urldef\tempurl%
\url{https://www.nytimes.com/2018/06/21/opinion/facial-analysis-technology-bias.html}
\showURL{%
\tempurl}


\bibitem[\protect\citeauthoryear{Callahan, Freire, Santos, Scheidegger, Silva,
  and Vo}{Callahan et~al\mbox{.}}{2006}]%
        {callahan2006vistrails}
\bibfield{author}{\bibinfo{person}{Steven~P Callahan}, \bibinfo{person}{Juliana
  Freire}, \bibinfo{person}{Emanuele Santos}, \bibinfo{person}{Carlos~E
  Scheidegger}, \bibinfo{person}{Cl{\'a}udio~T Silva}, {and}
  \bibinfo{person}{Huy~T Vo}.} \bibinfo{year}{2006}\natexlab{}.
\newblock \showarticletitle{VisTrails: visualization meets data management}. In
  \bibinfo{booktitle}{\emph{Proceedings of the 2006 ACM SIGMOD international
  conference on Management of data}}. ACM, \bibinfo{pages}{745--747}.
\newblock


\bibitem[\protect\citeauthoryear{Ceglowski}{Ceglowski}{2014}]%
        {ceglowski2014internet}
\bibfield{author}{\bibinfo{person}{Maciej Ceglowski}.}
  \bibinfo{year}{2014}\natexlab{}.
\newblock \showarticletitle{The Internet with a Human Face}.
\newblock \bibinfo{journal}{\emph{talk at Beyond Tellerrand in D{\"u}sseldorf,
  Germany, on}}  \bibinfo{volume}{20} (\bibinfo{year}{2014}).
\newblock


\bibitem[\protect\citeauthoryear{Chang, Gerrish, Wang, Boyd-Graber, and
  Blei}{Chang et~al\mbox{.}}{2009}]%
        {chang2009reading}
\bibfield{author}{\bibinfo{person}{Jonathan Chang}, \bibinfo{person}{Sean
  Gerrish}, \bibinfo{person}{Chong Wang}, \bibinfo{person}{Jordan~L
  Boyd-Graber}, {and} \bibinfo{person}{David~M Blei}.}
  \bibinfo{year}{2009}\natexlab{}.
\newblock \showarticletitle{Reading tea leaves: How humans interpret topic
  models}. In \bibinfo{booktitle}{\emph{Advances in neural information
  processing systems}}. \bibinfo{pages}{288--296}.
\newblock


\bibitem[\protect\citeauthoryear{Citron}{Citron}{2007}]%
        {citron2007technological}
\bibfield{author}{\bibinfo{person}{Danielle~Keats Citron}.}
  \bibinfo{year}{2007}\natexlab{}.
\newblock \showarticletitle{Technological due process}.
\newblock \bibinfo{journal}{\emph{Wash. UL Rev.}}  \bibinfo{volume}{85}
  (\bibinfo{year}{2007}), \bibinfo{pages}{1249}.
\newblock


\bibitem[\protect\citeauthoryear{Correll and Gleicher}{Correll and
  Gleicher}{2014}]%
        {correll2014error}
\bibfield{author}{\bibinfo{person}{Michael Correll} {and}
  \bibinfo{person}{Michael Gleicher}.} \bibinfo{year}{2014}\natexlab{}.
\newblock \showarticletitle{Error bars considered harmful: Exploring alternate
  encodings for mean and error}.
\newblock \bibinfo{journal}{\emph{IEEE transactions on visualization and
  computer graphics}} \bibinfo{volume}{20}, \bibinfo{number}{12}
  (\bibinfo{year}{2014}), \bibinfo{pages}{2142--2151}.
\newblock


\bibitem[\protect\citeauthoryear{Correll and Heer}{Correll and Heer}{2017}]%
        {correll2017black}
\bibfield{author}{\bibinfo{person}{Michael Correll} {and}
  \bibinfo{person}{Jeffrey Heer}.} \bibinfo{year}{2017}\natexlab{}.
\newblock \showarticletitle{Black hat visualization}. In
  \bibinfo{booktitle}{\emph{Workshop on Dealing with Cognitive Biases in
  Visualisations (DECISIVe), IEEE VIS}}.
\newblock


\bibitem[\protect\citeauthoryear{Crampton}{Crampton}{2011}]%
        {crampton2011mapping}
\bibfield{author}{\bibinfo{person}{Jeremy~W Crampton}.}
  \bibinfo{year}{2011}\natexlab{}.
\newblock \bibinfo{booktitle}{\emph{Mapping: A critical introduction to
  cartography and GIS}}. Vol.~\bibinfo{volume}{11}.
\newblock \bibinfo{publisher}{John Wiley \& Sons}.
\newblock


\bibitem[\protect\citeauthoryear{Crampton and Krygier}{Crampton and
  Krygier}{2006}]%
        {crampton2006introduction}
\bibfield{author}{\bibinfo{person}{Jeremy~W Crampton} {and}
  \bibinfo{person}{John Krygier}.} \bibinfo{year}{2006}\natexlab{}.
\newblock \showarticletitle{An introduction to critical cartography}.
\newblock \bibinfo{journal}{\emph{ACME: an International E-journal for Critical
  Geographies}} \bibinfo{volume}{4}, \bibinfo{number}{1}
  (\bibinfo{year}{2006}), \bibinfo{pages}{11--33}.
\newblock


\bibitem[\protect\citeauthoryear{Dalton and Thatcher}{Dalton and
  Thatcher}{2014}]%
        {dalton2014does}
\bibfield{author}{\bibinfo{person}{Craig Dalton} {and} \bibinfo{person}{Jim
  Thatcher}.} \bibinfo{year}{2014}\natexlab{}.
\newblock \showarticletitle{What does a critical data studies look like, and
  why do we care? Seven points for a critical approach to `big data'}.
\newblock \bibinfo{journal}{\emph{Society and Space}}  \bibinfo{volume}{29}
  (\bibinfo{year}{2014}).
\newblock


\bibitem[\protect\citeauthoryear{Dasgupta and Kosara}{Dasgupta and
  Kosara}{2011}]%
        {dasgupta2011adaptive}
\bibfield{author}{\bibinfo{person}{Aritra Dasgupta} {and}
  \bibinfo{person}{Robert Kosara}.} \bibinfo{year}{2011}\natexlab{}.
\newblock \showarticletitle{Adaptive privacy-preserving visualization using
  parallel coordinates}.
\newblock \bibinfo{journal}{\emph{IEEE Transactions on Visualization and
  Computer Graphics}} \bibinfo{volume}{17}, \bibinfo{number}{12}
  (\bibinfo{year}{2011}), \bibinfo{pages}{2241--2248}.
\newblock


\bibitem[\protect\citeauthoryear{Day}{Day}{2017}]%
        {datalabor}
\bibfield{author}{\bibinfo{person}{Deanna Day}.}
  \bibinfo{year}{2017}\natexlab{}.
\newblock \showarticletitle{The History of Data is the History of Labor}.
\newblock \bibinfo{journal}{\emph{The New Inquiry}} (\bibinfo{date}{Mar}
  \bibinfo{year}{2017}).
\newblock
\urldef\tempurl%
\url{https://thenewinquiry.com/blog/the-history-of-data-is-the-history-of-labor/}
\showURL{%
\tempurl}


\bibitem[\protect\citeauthoryear{D'Ignazio and Klein}{D'Ignazio and Klein}{[n.
  d.]}]%
        {dignazio2019draft}
\bibfield{author}{\bibinfo{person}{Catherine D'Ignazio} {and}
  \bibinfo{person}{Lauren Klein}.} \bibinfo{year}{[n. d.]}\natexlab{}.
\newblock \bibinfo{booktitle}{\emph{Data Feminism}}.
\newblock
\newblock
\shownote{2018 Draft.}


\bibitem[\protect\citeauthoryear{D'Ignazio and Klein}{D'Ignazio and
  Klein}{2016}]%
        {d2016feminist}
\bibfield{author}{\bibinfo{person}{Catherine D'Ignazio} {and}
  \bibinfo{person}{Lauren~F Klein}.} \bibinfo{year}{2016}\natexlab{}.
\newblock \showarticletitle{Feminist data visualization}. In
  \bibinfo{booktitle}{\emph{Workshop on Visualization for the Digital
  Humanities (VIS4DH), Baltimore. IEEE}}.
\newblock


\bibitem[\protect\citeauthoryear{Dillard}{Dillard}{2003}]%
        {dillard2003professional}
\bibfield{author}{\bibinfo{person}{Jesse~F Dillard}.}
  \bibinfo{year}{2003}\natexlab{}.
\newblock \showarticletitle{Professional services, IBM, and the Holocaust}.
\newblock \bibinfo{journal}{\emph{Journal of Information Systems}}
  \bibinfo{volume}{17}, \bibinfo{number}{2} (\bibinfo{year}{2003}),
  \bibinfo{pages}{1--16}.
\newblock


\bibitem[\protect\citeauthoryear{Dimara, Bailly, Bezerianos, and
  Franconeri}{Dimara et~al\mbox{.}}{2018}]%
        {dimara2018}
\bibfield{author}{\bibinfo{person}{E. Dimara}, \bibinfo{person}{G. Bailly},
  \bibinfo{person}{A. Bezerianos}, {and} \bibinfo{person}{S. Franconeri}.}
  \bibinfo{year}{2018}\natexlab{}.
\newblock \showarticletitle{Mitigating the Attraction Effect with
  Visualizations}.
\newblock \bibinfo{journal}{\emph{IEEE Transactions on Visualization and
  Computer Graphics}} (\bibinfo{year}{2018}), \bibinfo{pages}{1--1}.
\newblock
\showISSN{1077-2626}
\urldef\tempurl%
\url{https://doi.org/10.1109/TVCG.2018.2865233}
\showDOI{\tempurl}


\bibitem[\protect\citeauthoryear{D{\"o}rk, Feng, Collins, and
  Carpendale}{D{\"o}rk et~al\mbox{.}}{2013}]%
        {dork2013critical}
\bibfield{author}{\bibinfo{person}{Marian D{\"o}rk}, \bibinfo{person}{Patrick
  Feng}, \bibinfo{person}{Christopher Collins}, {and} \bibinfo{person}{Sheelagh
  Carpendale}.} \bibinfo{year}{2013}\natexlab{}.
\newblock \showarticletitle{Critical InfoVis: exploring the politics of
  visualization}. In \bibinfo{booktitle}{\emph{CHI'13 Extended Abstracts on
  Human Factors in Computing Systems}}. ACM, \bibinfo{pages}{2189--2198}.
\newblock


\bibitem[\protect\citeauthoryear{Dragga and Voss}{Dragga and Voss}{2001}]%
        {dragga2001cruel}
\bibfield{author}{\bibinfo{person}{Sam Dragga} {and} \bibinfo{person}{Dan
  Voss}.} \bibinfo{year}{2001}\natexlab{}.
\newblock \showarticletitle{Cruel pies: The inhumanity of technical
  illustrations}.
\newblock \bibinfo{journal}{\emph{Technical communication}}
  \bibinfo{volume}{48}, \bibinfo{number}{3} (\bibinfo{year}{2001}),
  \bibinfo{pages}{265--274}.
\newblock


\bibitem[\protect\citeauthoryear{Drucker}{Drucker}{2012}]%
        {drucker2012humanistic}
\bibfield{author}{\bibinfo{person}{Johanna Drucker}.}
  \bibinfo{year}{2012}\natexlab{}.
\newblock \showarticletitle{Humanistic theory and digital scholarship}.
\newblock \bibinfo{journal}{\emph{Debates in the digital humanities}}
  (\bibinfo{year}{2012}), \bibinfo{pages}{85--95}.
\newblock


\bibitem[\protect\citeauthoryear{Du~Bois}{Du~Bois}{1900}]%
        {dubois1900}
\bibfield{author}{\bibinfo{person}{W.E.B. Du~Bois}.}
  \bibinfo{year}{1900}\natexlab{}.
\newblock \bibinfo{title}{Proportion of freemen and slaves among American
  Negroes}.
\newblock
  \bibinfo{howpublished}{\url{http://hdl.loc.gov/loc.pnp/ppmsca.33913}}.
\newblock
\newblock
\shownote{A series of statistical charts illustrating the condition of the
  descendants of former African slaves now in residence in the United States of
  America.}


\bibitem[\protect\citeauthoryear{Du~Bois et~al\mbox{.}}{Du~Bois
  et~al\mbox{.}}{1926}]%
        {du1926criteria}
\bibfield{author}{\bibinfo{person}{W.E.B. Du~Bois} {et~al\mbox{.}}}
  \bibinfo{year}{1926}\natexlab{}.
\newblock \showarticletitle{Criteria of Negro art}.
\newblock \bibinfo{journal}{\emph{Crisis}} \bibinfo{volume}{32},
  \bibinfo{number}{6} (\bibinfo{year}{1926}), \bibinfo{pages}{290--297}.
\newblock


\bibitem[\protect\citeauthoryear{Farley, Grady, Miller, O'Connor, Schneider,
  Spikes, Constantinou, et~al\mbox{.}}{Farley et~al\mbox{.}}{2014}]%
        {farley2014spj}
\bibfield{author}{\bibinfo{person}{Elizabeth Farley}, \bibinfo{person}{Fiona
  Grady}, \bibinfo{person}{Dean~S Miller}, \bibinfo{person}{Rory O'Connor},
  \bibinfo{person}{Howard Schneider}, \bibinfo{person}{Michael Spikes},
  \bibinfo{person}{Constantia Constantinou}, {et~al\mbox{.}}}
  \bibinfo{year}{2014}\natexlab{}.
\newblock \showarticletitle{SPJ Code of Ethics}.
\newblock \bibinfo{journal}{\emph{The Power of Images}} (\bibinfo{year}{2014}).
\newblock


\bibitem[\protect\citeauthoryear{Feng, Deng, Peck, and Harrison}{Feng
  et~al\mbox{.}}{2017}]%
        {feng2017hindsight}
\bibfield{author}{\bibinfo{person}{Mi Feng}, \bibinfo{person}{Cheng Deng},
  \bibinfo{person}{Evan~M Peck}, {and} \bibinfo{person}{Lane Harrison}.}
  \bibinfo{year}{2017}\natexlab{}.
\newblock \showarticletitle{HindSight: Encouraging exploration through direct
  encoding of personal interaction history}.
\newblock \bibinfo{journal}{\emph{IEEE transactions on visualization and
  computer graphics}} \bibinfo{volume}{23}, \bibinfo{number}{1}
  (\bibinfo{year}{2017}), \bibinfo{pages}{351--360}.
\newblock


\bibitem[\protect\citeauthoryear{Gallagher}{Gallagher}{2018a}]%
        {gallagher2018b}
\bibfield{author}{\bibinfo{person}{Ryan Gallagher}.}
  \bibinfo{year}{2018}\natexlab{a}.
\newblock \showarticletitle{Google Shut Out Privacy and Security Teams From
  Secret China Project}.
\newblock \bibinfo{journal}{\emph{The Intercept}} (\bibinfo{date}{Nov}
  \bibinfo{year}{2018}).
\newblock
\urldef\tempurl%
\url{https://theintercept.com/2018/11/29/google-china-censored-search/}
\showURL{%
\tempurl}


\bibitem[\protect\citeauthoryear{Gallagher}{Gallagher}{2018b}]%
        {gallagher2018}
\bibfield{author}{\bibinfo{person}{Ryan Gallagher}.}
  \bibinfo{year}{2018}\natexlab{b}.
\newblock \showarticletitle{Senior Google Scientist Resigns Over ``Foreiture of
  our Values'' in China}.
\newblock \bibinfo{journal}{\emph{The Intercept}} (\bibinfo{date}{Sep}
  \bibinfo{year}{2018}).
\newblock
\urldef\tempurl%
\url{https://theintercept.com/2018/09/13/google-china-search-engine-employee-resigns/}
\showURL{%
\tempurl}


\bibitem[\protect\citeauthoryear{Gelman and Loken}{Gelman and Loken}{2013}]%
        {gelman2013garden}
\bibfield{author}{\bibinfo{person}{Andrew Gelman} {and} \bibinfo{person}{Eric
  Loken}.} \bibinfo{year}{2013}\natexlab{}.
\newblock \showarticletitle{The garden of forking paths: Why multiple
  comparisons can be a problem, even when there is no ``fishing expedition'' or
  ``p-hacking'' and the research hypothesis was posited ahead of time}.
\newblock \bibinfo{journal}{\emph{Department of Statistics, Columbia
  University}} (\bibinfo{year}{2013}).
\newblock


\bibitem[\protect\citeauthoryear{Gitelman}{Gitelman}{2013}]%
        {gitelman2013raw}
\bibfield{author}{\bibinfo{person}{Lisa Gitelman}.}
  \bibinfo{year}{2013}\natexlab{}.
\newblock \bibinfo{booktitle}{\emph{Raw data is an oxymoron}}.
\newblock \bibinfo{publisher}{MIT Press}.
\newblock


\bibitem[\protect\citeauthoryear{Gleicher}{Gleicher}{2013}]%
        {gleicher2013explainers}
\bibfield{author}{\bibinfo{person}{Michael Gleicher}.}
  \bibinfo{year}{2013}\natexlab{}.
\newblock \showarticletitle{Explainers: Expert explorations with crafted
  projections}.
\newblock \bibinfo{journal}{\emph{IEEE Transactions on Visualization \&
  Computer Graphics}} \bibinfo{number}{12} (\bibinfo{year}{2013}),
  \bibinfo{pages}{2042--2051}.
\newblock


\bibitem[\protect\citeauthoryear{Greis, Hullman, Correll, Kay, and Shaer}{Greis
  et~al\mbox{.}}{2017}]%
        {greis2017designing}
\bibfield{author}{\bibinfo{person}{Miriam Greis}, \bibinfo{person}{Jessica
  Hullman}, \bibinfo{person}{Michael Correll}, \bibinfo{person}{Matthew Kay},
  {and} \bibinfo{person}{Orit Shaer}.} \bibinfo{year}{2017}\natexlab{}.
\newblock \showarticletitle{Designing for Uncertainty in HCI: When Does
  Uncertainty Help?}. In \bibinfo{booktitle}{\emph{Proceedings of the 2017 CHI
  Conference Extended Abstracts on Human Factors in Computing Systems}}. ACM,
  \bibinfo{pages}{593--600}.
\newblock


\bibitem[\protect\citeauthoryear{Haraway}{Haraway}{1988}]%
        {haraway1988situated}
\bibfield{author}{\bibinfo{person}{Donna Haraway}.}
  \bibinfo{year}{1988}\natexlab{}.
\newblock \showarticletitle{Situated knowledges: The science question in
  feminism and the privilege of partial perspective}.
\newblock \bibinfo{journal}{\emph{Feminist studies}} \bibinfo{volume}{14},
  \bibinfo{number}{3} (\bibinfo{year}{1988}), \bibinfo{pages}{575--599}.
\newblock


\bibitem[\protect\citeauthoryear{Hecht, Wilcox, Bigham, Schöning, Hoque,
  Ernst, Bisk, De~Russis, Yarosh, Anjum, Contractor, and Wu}{Hecht
  et~al\mbox{.}}{2018}]%
        {fca2018}
\bibfield{author}{\bibinfo{person}{B. Hecht}, \bibinfo{person}{L. Wilcox},
  \bibinfo{person}{J.P. Bigham}, \bibinfo{person}{J. Schöning},
  \bibinfo{person}{E. Hoque}, \bibinfo{person}{J. Ernst}, \bibinfo{person}{Y.
  Bisk}, \bibinfo{person}{L. De~Russis}, \bibinfo{person}{L. Yarosh},
  \bibinfo{person}{B. Anjum}, \bibinfo{person}{D. Contractor}, {and}
  \bibinfo{person}{C. Wu}.} \bibinfo{year}{2018}\natexlab{}.
\newblock \showarticletitle{It's Time to Do Something: Mitigating the Negative
  Impacts of Computing Through a Change to the Peer Review Process}.
\newblock \bibinfo{journal}{\emph{ACM Future of Computing Blog}}
  (\bibinfo{date}{Mar} \bibinfo{year}{2018}).
\newblock
\urldef\tempurl%
\url{https://acm-fca.org/2018/03/29/negativeimpacts/}
\showURL{%
\tempurl}


\bibitem[\protect\citeauthoryear{Heidegger}{Heidegger}{1954}]%
        {heidegger1954question}
\bibfield{author}{\bibinfo{person}{Martin Heidegger}.}
  \bibinfo{year}{1954}\natexlab{}.
\newblock \showarticletitle{The question concerning technology}.
\newblock \bibinfo{journal}{\emph{Technology and values: Essential readings}}
  \bibinfo{volume}{99} (\bibinfo{year}{1954}), \bibinfo{pages}{113}.
\newblock


\bibitem[\protect\citeauthoryear{Henrich, Heine, and Norenzayan}{Henrich
  et~al\mbox{.}}{2010}]%
        {henrich2010most}
\bibfield{author}{\bibinfo{person}{Joseph Henrich}, \bibinfo{person}{Steven~J
  Heine}, {and} \bibinfo{person}{Ara Norenzayan}.}
  \bibinfo{year}{2010}\natexlab{}.
\newblock \showarticletitle{Most people are not WEIRD}.
\newblock \bibinfo{journal}{\emph{Nature}} \bibinfo{volume}{466},
  \bibinfo{number}{7302} (\bibinfo{year}{2010}), \bibinfo{pages}{29}.
\newblock


\bibitem[\protect\citeauthoryear{Hill, Kennedy, and Gerrard}{Hill
  et~al\mbox{.}}{2016}]%
        {hill2016visualizing}
\bibfield{author}{\bibinfo{person}{Rosemary~Lucy Hill}, \bibinfo{person}{Helen
  Kennedy}, {and} \bibinfo{person}{Ysabel Gerrard}.}
  \bibinfo{year}{2016}\natexlab{}.
\newblock \showarticletitle{Visualizing junk: Big data visualizations and the
  need for feminist data studies}.
\newblock \bibinfo{journal}{\emph{Journal of Communication Inquiry}}
  \bibinfo{volume}{40}, \bibinfo{number}{4} (\bibinfo{year}{2016}),
  \bibinfo{pages}{331--350}.
\newblock


\bibitem[\protect\citeauthoryear{Hochshild}{Hochshild}{1983}]%
        {hochshild1983managed}
\bibfield{author}{\bibinfo{person}{Arlie~Russell Hochshild}.}
  \bibinfo{year}{1983}\natexlab{}.
\newblock \bibinfo{booktitle}{\emph{The Managed Heart: Commercialization of
  Human Feeling}}.
\newblock \bibinfo{publisher}{The University of California Press}.
\newblock


\bibitem[\protect\citeauthoryear{Hsu}{Hsu}{2018}]%
        {hsu2018}
\bibfield{author}{\bibinfo{person}{Jeremy Hsu}.}
  \bibinfo{year}{2018}\natexlab{}.
\newblock \showarticletitle{The Strava Heat Map and the End of Secrets}.
\newblock \bibinfo{journal}{\emph{Wired}} (\bibinfo{date}{Jan}
  \bibinfo{year}{2018}).
\newblock
\urldef\tempurl%
\url{https://www.wired.com/story/strava-heat-map-military-bases-fitness-trackers-privacy/}
\showURL{%
\tempurl}


\bibitem[\protect\citeauthoryear{Hullman and Diakopoulos}{Hullman and
  Diakopoulos}{2011}]%
        {hullman2011visualization}
\bibfield{author}{\bibinfo{person}{Jessica Hullman} {and} \bibinfo{person}{Nick
  Diakopoulos}.} \bibinfo{year}{2011}\natexlab{}.
\newblock \showarticletitle{Visualization rhetoric: Framing effects in
  narrative visualization}.
\newblock \bibinfo{journal}{\emph{IEEE transactions on visualization and
  computer graphics}} \bibinfo{volume}{17}, \bibinfo{number}{12}
  (\bibinfo{year}{2011}), \bibinfo{pages}{2231--2240}.
\newblock


\bibitem[\protect\citeauthoryear{Hullman, Qiao, Correll, Kale, and Kay}{Hullman
  et~al\mbox{.}}{2018}]%
        {hullman2018pursuit}
\bibfield{author}{\bibinfo{person}{Jessica Hullman}, \bibinfo{person}{Xiaoli
  Qiao}, \bibinfo{person}{Michael Correll}, \bibinfo{person}{Alex Kale}, {and}
  \bibinfo{person}{Matthew Kay}.} \bibinfo{year}{2018}\natexlab{}.
\newblock \showarticletitle{In Pursuit of Error: A Survey of Uncertainty
  Visualization Evaluation}.
\newblock \bibinfo{journal}{\emph{IEEE transactions on visualization and
  computer graphics}} (\bibinfo{year}{2018}).
\newblock


\bibitem[\protect\citeauthoryear{Hursthouse}{Hursthouse}{1999}]%
        {hursthouse1999virtue}
\bibfield{author}{\bibinfo{person}{Rosalind Hursthouse}.}
  \bibinfo{year}{1999}\natexlab{}.
\newblock \bibinfo{booktitle}{\emph{On virtue ethics}}.
\newblock \bibinfo{publisher}{OUP Oxford}.
\newblock


\bibitem[\protect\citeauthoryear{Jeong}{Jeong}{2018}]%
        {jeong2018}
\bibfield{author}{\bibinfo{person}{Sarah Jeong}.}
  \bibinfo{year}{2018}\natexlab{}.
\newblock \showarticletitle{Bad Romance: How a cabal of authors profited by
  gaming Amazon's Kindle Unlimited algorithm}.
\newblock \bibinfo{journal}{\emph{The Verge}} (\bibinfo{date}{Jul}
  \bibinfo{year}{2018}).
\newblock
\urldef\tempurl%
\url{https://www.theverge.com/2018/7/16/17566276/}
\showURL{%
\tempurl}


\bibitem[\protect\citeauthoryear{Kay, Kola, Hullman, and Munson}{Kay
  et~al\mbox{.}}{2016}]%
        {kay2016ish}
\bibfield{author}{\bibinfo{person}{Matthew Kay}, \bibinfo{person}{Tara Kola},
  \bibinfo{person}{Jessica~R Hullman}, {and} \bibinfo{person}{Sean~A Munson}.}
  \bibinfo{year}{2016}\natexlab{}.
\newblock \showarticletitle{When (ish) is my bus?: User-centered visualizations
  of uncertainty in everyday, mobile predictive systems}. In
  \bibinfo{booktitle}{\emph{Proceedings of the 2016 CHI Conference on Human
  Factors in Computing Systems}}. ACM, \bibinfo{pages}{5092--5103}.
\newblock


\bibitem[\protect\citeauthoryear{Kennedy, Hill, Aiello, and Allen}{Kennedy
  et~al\mbox{.}}{2016}]%
        {kennedy2016work}
\bibfield{author}{\bibinfo{person}{Helen Kennedy},
  \bibinfo{person}{Rosemary~Lucy Hill}, \bibinfo{person}{Giorgia Aiello}, {and}
  \bibinfo{person}{William Allen}.} \bibinfo{year}{2016}\natexlab{}.
\newblock \showarticletitle{The work that visualisation conventions do}.
\newblock \bibinfo{journal}{\emph{Information, Communication \& Society}}
  \bibinfo{volume}{19}, \bibinfo{number}{6} (\bibinfo{year}{2016}),
  \bibinfo{pages}{715--735}.
\newblock


\bibitem[\protect\citeauthoryear{Kim, Khanna, and Koyejo}{Kim
  et~al\mbox{.}}{2016}]%
        {kim2016MMD}
\bibfield{author}{\bibinfo{person}{Been Kim}, \bibinfo{person}{Rajiv Khanna},
  {and} \bibinfo{person}{Sanmi Koyejo}.} \bibinfo{year}{2016}\natexlab{}.
\newblock \showarticletitle{Examples are not Enough, Learn to Criticize!
  {C}riticism for Interpretability}. In \bibinfo{booktitle}{\emph{Advances in
  Neural Information Processing Systems}}.
\newblock


\bibitem[\protect\citeauthoryear{{Kindermans}, {Hooker}, {Adebayo}, {Alber},
  {Sch{\"u}tt}, {D{\"a}hne}, {Erhan}, and {Kim}}{{Kindermans}
  et~al\mbox{.}}{2017}]%
        {2017arXiv171100867K}
\bibfield{author}{\bibinfo{person}{P.-J. {Kindermans}}, \bibinfo{person}{S.
  {Hooker}}, \bibinfo{person}{J. {Adebayo}}, \bibinfo{person}{M. {Alber}},
  \bibinfo{person}{K.~T. {Sch{\"u}tt}}, \bibinfo{person}{S. {D{\"a}hne}},
  \bibinfo{person}{D. {Erhan}}, {and} \bibinfo{person}{B. {Kim}}.}
  \bibinfo{year}{2017}\natexlab{}.
\newblock \showarticletitle{{The (Un)reliability of saliency methods}}.
\newblock \bibinfo{journal}{\emph{NIPS workshop on Explaining and Visualizing
  Deep Learning}} (\bibinfo{year}{2017}).
\newblock
\showeprint{stat.ML/1711.00867}


\bibitem[\protect\citeauthoryear{Kong, Liu, and Karahalios}{Kong
  et~al\mbox{.}}{2018}]%
        {kong2018frames}
\bibfield{author}{\bibinfo{person}{Ha-Kyung Kong}, \bibinfo{person}{Zhicheng
  Liu}, {and} \bibinfo{person}{Karrie Karahalios}.}
  \bibinfo{year}{2018}\natexlab{}.
\newblock \showarticletitle{Frames and Slants in Titles of Visualizations on
  Controversial Topics}. In \bibinfo{booktitle}{\emph{Proceedings of the 2018
  CHI Conference on Human Factors in Computing Systems}}. ACM,
  \bibinfo{pages}{438}.
\newblock


\bibitem[\protect\citeauthoryear{Kosara}{Kosara}{2018}]%
        {kosara2018}
\bibfield{author}{\bibinfo{person}{Robert Kosara}.}
  \bibinfo{year}{2018}\natexlab{}.
\newblock \bibinfo{title}{How to Get Excited About Standard Datasets}.
\newblock
  \bibinfo{howpublished}{\url{https://eagereyes.org/blog/2018/how-to-get-excited-about-standard-datasets}}.
\newblock


\bibitem[\protect\citeauthoryear{Kosara and Haroz}{Kosara and Haroz}{2018}]%
        {kosara2018skipping}
\bibfield{author}{\bibinfo{person}{Robert Kosara} {and} \bibinfo{person}{Steve
  Haroz}.} \bibinfo{year}{2018}\natexlab{}.
\newblock \showarticletitle{Skipping the Replication Crisis in Visualization:
  Threats to Study Validity and How to Address Them}. In
  \bibinfo{booktitle}{\emph{Proceedings of BELIV 2018: Evaluation and Beyond --
  Methodological Approaches for Visualization}}.
\newblock


\bibitem[\protect\citeauthoryear{Kosara and Mackinlay}{Kosara and
  Mackinlay}{2013}]%
        {kosara2013storytelling}
\bibfield{author}{\bibinfo{person}{Robert Kosara} {and} \bibinfo{person}{Jock
  Mackinlay}.} \bibinfo{year}{2013}\natexlab{}.
\newblock \showarticletitle{Storytelling: The next step for visualization}.
\newblock \bibinfo{journal}{\emph{Computer}} \bibinfo{volume}{46},
  \bibinfo{number}{5} (\bibinfo{year}{2013}), \bibinfo{pages}{44--50}.
\newblock


\bibitem[\protect\citeauthoryear{Lecher}{Lecher}{2018}]%
        {lecher2018}
\bibfield{author}{\bibinfo{person}{Colin Lecher}.}
  \bibinfo{year}{2018}\natexlab{}.
\newblock \showarticletitle{What Happens When An Algorithm Cuts Your Health
  Care}.
\newblock \bibinfo{journal}{\emph{The Verge}} (\bibinfo{date}{Mar}
  \bibinfo{year}{2018}).
\newblock
\urldef\tempurl%
\url{https://www.theverge.com/2018/3/21/17144260/}
\showURL{%
\tempurl}


\bibitem[\protect\citeauthoryear{Lewis and Willis}{Lewis and Willis}{2010}]%
        {lewis2010small}
\bibfield{author}{\bibinfo{person}{David~Levering Lewis} {and}
  \bibinfo{person}{Deborah Willis}.} \bibinfo{year}{2010}\natexlab{}.
\newblock \bibinfo{booktitle}{\emph{A small nation of people: W.E.B. Du Bois
  and African American portraits of progress}}.
\newblock \bibinfo{publisher}{Zondervan}.
\newblock


\bibitem[\protect\citeauthoryear{Lipton}{Lipton}{2016}]%
        {lipton2016mythos}
\bibfield{author}{\bibinfo{person}{Zachary~C Lipton}.}
  \bibinfo{year}{2016}\natexlab{}.
\newblock \showarticletitle{The mythos of model interpretability}.
\newblock \bibinfo{journal}{\emph{arXiv preprint arXiv:1606.03490}}
  (\bibinfo{year}{2016}).
\newblock


\bibitem[\protect\citeauthoryear{Louden}{Louden}{1984}]%
        {louden1984some}
\bibfield{author}{\bibinfo{person}{Robert~B Louden}.}
  \bibinfo{year}{1984}\natexlab{}.
\newblock \showarticletitle{On some vices of virtue ethics}.
\newblock \bibinfo{journal}{\emph{American Philosophical Quarterly}}
  \bibinfo{volume}{21}, \bibinfo{number}{3} (\bibinfo{year}{1984}),
  \bibinfo{pages}{227--236}.
\newblock


\bibitem[\protect\citeauthoryear{Lundberg and Lee}{Lundberg and Lee}{2017}]%
        {lundberg2017unified}
\bibfield{author}{\bibinfo{person}{Scott~M Lundberg} {and}
  \bibinfo{person}{Su-In Lee}.} \bibinfo{year}{2017}\natexlab{}.
\newblock \showarticletitle{A unified approach to interpreting model
  predictions}. In \bibinfo{booktitle}{\emph{Advances in Neural Information
  Processing Systems}}. \bibinfo{pages}{4765--4774}.
\newblock


\bibitem[\protect\citeauthoryear{Meyer and Blohm}{Meyer and Blohm}{1942}]%
        {meyer1942landvolk}
\bibfield{author}{\bibinfo{person}{Konrad Meyer} {and} \bibinfo{person}{Georg
  Blohm}.} \bibinfo{year}{1942}\natexlab{}.
\newblock \bibinfo{booktitle}{\emph{Landvolk im Werden}}.
\newblock \bibinfo{publisher}{Deutsche Landbuchhandlung}.
\newblock


\bibitem[\protect\citeauthoryear{Micallef, Dragicevic, and Fekete}{Micallef
  et~al\mbox{.}}{2012}]%
        {micallef2012assessing}
\bibfield{author}{\bibinfo{person}{Luana Micallef}, \bibinfo{person}{Pierre
  Dragicevic}, {and} \bibinfo{person}{Jean-Daniel Fekete}.}
  \bibinfo{year}{2012}\natexlab{}.
\newblock \showarticletitle{Assessing the effect of visualizations on bayesian
  reasoning through crowdsourcing}.
\newblock \bibinfo{journal}{\emph{IEEE Transactions on Visualization and
  Computer Graphics}} \bibinfo{volume}{18}, \bibinfo{number}{12}
  (\bibinfo{year}{2012}), \bibinfo{pages}{2536--2545}.
\newblock


\bibitem[\protect\citeauthoryear{Muehlenhaus}{Muehlenhaus}{2013}]%
        {muehlenhaus2013design}
\bibfield{author}{\bibinfo{person}{Ian Muehlenhaus}.}
  \bibinfo{year}{2013}\natexlab{}.
\newblock \showarticletitle{The design and composition of persuasive maps}.
\newblock \bibinfo{journal}{\emph{Cartography and Geographic Information
  Science}} \bibinfo{volume}{40}, \bibinfo{number}{5} (\bibinfo{year}{2013}),
  \bibinfo{pages}{401--414}.
\newblock


\bibitem[\protect\citeauthoryear{Nichols}{Nichols}{2017}]%
        {nichols2017death}
\bibfield{author}{\bibinfo{person}{Thomas~M Nichols}.}
  \bibinfo{year}{2017}\natexlab{}.
\newblock \bibinfo{booktitle}{\emph{The death of expertise}}.
\newblock \bibinfo{publisher}{Tantor Media, Incorporated}.
\newblock


\bibitem[\protect\citeauthoryear{North}{North}{2006}]%
        {north2006toward}
\bibfield{author}{\bibinfo{person}{Chris North}.}
  \bibinfo{year}{2006}\natexlab{}.
\newblock \showarticletitle{Toward measuring visualization insight}.
\newblock \bibinfo{journal}{\emph{IEEE computer graphics and applications}}
  \bibinfo{volume}{26}, \bibinfo{number}{3} (\bibinfo{year}{2006}),
  \bibinfo{pages}{6--9}.
\newblock


\bibitem[\protect\citeauthoryear{O'Neil}{O'Neil}{2016}]%
        {o2016weapons}
\bibfield{author}{\bibinfo{person}{Cathy O'Neil}.}
  \bibinfo{year}{2016}\natexlab{}.
\newblock \bibinfo{booktitle}{\emph{Weapons of math destruction: How big data
  increases inequality and threatens democracy}}.
\newblock \bibinfo{publisher}{Broadway Books}.
\newblock


\bibitem[\protect\citeauthoryear{Paine}{Paine}{1776}]%
        {paine2004common}
\bibfield{author}{\bibinfo{person}{Thomas Paine}.}
  \bibinfo{year}{1776}\natexlab{}.
\newblock \bibinfo{booktitle}{\emph{Common sense}}.
\newblock


\bibitem[\protect\citeauthoryear{Pandey, Manivannan, Nov, Satterthwaite, and
  Bertini}{Pandey et~al\mbox{.}}{2014}]%
        {pandey2014persuasive}
\bibfield{author}{\bibinfo{person}{Anshul~Vikram Pandey},
  \bibinfo{person}{Anjali Manivannan}, \bibinfo{person}{Oded Nov},
  \bibinfo{person}{Margaret Satterthwaite}, {and} \bibinfo{person}{Enrico
  Bertini}.} \bibinfo{year}{2014}\natexlab{}.
\newblock \showarticletitle{The persuasive power of data visualization}.
\newblock \bibinfo{journal}{\emph{IEEE transactions on visualization and
  computer graphics}} \bibinfo{volume}{20}, \bibinfo{number}{12}
  (\bibinfo{year}{2014}), \bibinfo{pages}{2211--2220}.
\newblock


\bibitem[\protect\citeauthoryear{Patil, Mason, and Loukides}{Patil
  et~al\mbox{.}}{2018}]%
        {patil2018ethics}
\bibfield{author}{\bibinfo{person}{DJ Patil}, \bibinfo{person}{Hilary Mason},
  {and} \bibinfo{person}{Mike Loukides}.} \bibinfo{year}{2018}\natexlab{}.
\newblock \bibinfo{booktitle}{\emph{Ethics and Data Science}}.
\newblock \bibinfo{publisher}{O'Reilly}.
\newblock


\bibitem[\protect\citeauthoryear{Poulson}{Poulson}{2018}]%
        {poulson2018}
\bibfield{author}{\bibinfo{person}{Jack Poulson}.}
  \bibinfo{year}{2018}\natexlab{}.
\newblock \showarticletitle{I Quit Google Over Its Censored Chinese Search
  Engine. The Company Needs to Clarify Its Position on Human Rights.}
\newblock \bibinfo{journal}{\emph{The Intercept}} (\bibinfo{date}{Sep}
  \bibinfo{year}{2018}).
\newblock
\urldef\tempurl%
\url{https://theintercept.com/2018/12/01/google-china-censorship-human-rights/}
\showURL{%
\tempurl}


\bibitem[\protect\citeauthoryear{Pu and Kay}{Pu and Kay}{2018}]%
        {pugarden}
\bibfield{author}{\bibinfo{person}{Xiaoying Pu} {and} \bibinfo{person}{Matthew
  Kay}.} \bibinfo{year}{2018}\natexlab{}.
\newblock \showarticletitle{The garden of forking paths in visualization: A
  design space for reliable exploratory visual analytics}. In
  \bibinfo{booktitle}{\emph{Proceedings of the Workshop on Beyond Time and
  Errors: Novel Evaluation Methods for Visualization (BELIV)}}.
\newblock


\bibitem[\protect\citeauthoryear{Ribeiro, Singh, and Guestrin}{Ribeiro
  et~al\mbox{.}}{2016}]%
        {ribeiro2016should}
\bibfield{author}{\bibinfo{person}{Marco~Tulio Ribeiro},
  \bibinfo{person}{Sameer Singh}, {and} \bibinfo{person}{Carlos Guestrin}.}
  \bibinfo{year}{2016}\natexlab{}.
\newblock \showarticletitle{Why should i trust you?: Explaining the predictions
  of any classifier}. In \bibinfo{booktitle}{\emph{Proceedings of the 22nd ACM
  SIGKDD international conference on knowledge discovery and data mining}}.
  ACM, \bibinfo{pages}{1135--1144}.
\newblock


\bibitem[\protect\citeauthoryear{Richards}{Richards}{2003}]%
        {richards2003argument}
\bibfield{author}{\bibinfo{person}{Anne~R Richards}.}
  \bibinfo{year}{2003}\natexlab{}.
\newblock \showarticletitle{Argument and authority in the visual
  representations of science}.
\newblock \bibinfo{journal}{\emph{Technical Communication Quarterly}}
  \bibinfo{volume}{12}, \bibinfo{number}{2} (\bibinfo{year}{2003}),
  \bibinfo{pages}{183--206}.
\newblock


\bibitem[\protect\citeauthoryear{Rogaway}{Rogaway}{2015}]%
        {rogaway2015moral}
\bibfield{author}{\bibinfo{person}{Phillip Rogaway}.}
  \bibinfo{year}{2015}\natexlab{}.
\newblock \showarticletitle{The Moral Character of Cryptographic Work.}
\newblock \bibinfo{journal}{\emph{IACR Cryptology ePrint Archive}}
  \bibinfo{volume}{2015} (\bibinfo{year}{2015}), \bibinfo{pages}{1162}.
\newblock


\bibitem[\protect\citeauthoryear{Ruginski, Boone, Padilla, Liu, Heydari,
  Kramer, Hegarty, Thompson, House, and Creem-Regehr}{Ruginski
  et~al\mbox{.}}{2016}]%
        {ruginski2016non}
\bibfield{author}{\bibinfo{person}{Ian~T Ruginski},
  \bibinfo{person}{Alexander~P Boone}, \bibinfo{person}{Lace~M Padilla},
  \bibinfo{person}{Le Liu}, \bibinfo{person}{Nahal Heydari},
  \bibinfo{person}{Heidi~S Kramer}, \bibinfo{person}{Mary Hegarty},
  \bibinfo{person}{William~B Thompson}, \bibinfo{person}{Donald~H House}, {and}
  \bibinfo{person}{Sarah~H Creem-Regehr}.} \bibinfo{year}{2016}\natexlab{}.
\newblock \showarticletitle{Non-expert interpretations of hurricane forecast
  uncertainty visualizations}.
\newblock \bibinfo{journal}{\emph{Spatial Cognition \& Computation}}
  \bibinfo{volume}{16}, \bibinfo{number}{2} (\bibinfo{year}{2016}),
  \bibinfo{pages}{154--172}.
\newblock


\bibitem[\protect\citeauthoryear{Segel and Heer}{Segel and Heer}{2010}]%
        {segel2010narrative}
\bibfield{author}{\bibinfo{person}{Edward Segel} {and} \bibinfo{person}{Jeffrey
  Heer}.} \bibinfo{year}{2010}\natexlab{}.
\newblock \showarticletitle{Narrative visualization: Telling stories with
  data}.
\newblock \bibinfo{journal}{\emph{IEEE transactions on visualization and
  computer graphics}} \bibinfo{volume}{16}, \bibinfo{number}{6}
  (\bibinfo{year}{2010}), \bibinfo{pages}{1139--1148}.
\newblock


\bibitem[\protect\citeauthoryear{Shaban}{Shaban}{2018}]%
        {shaban2018}
\bibfield{author}{\bibinfo{person}{Hamza Shaban}.}
  \bibinfo{year}{2018}\natexlab{}.
\newblock \showarticletitle{Amazon employees demand company cut ties with ICE}.
\newblock \bibinfo{journal}{\emph{Washington Post}} (\bibinfo{date}{June}
  \bibinfo{year}{2018}).
\newblock
\urldef\tempurl%
\url{https://www.washingtonpost.com/news/the-switch/wp/2018/06/22/amazon-employees-demand-company-cut-ties-with-ice/}
\showURL{%
\tempurl}


\bibitem[\protect\citeauthoryear{Slobin}{Slobin}{2014}]%
        {slobin2014}
\bibfield{author}{\bibinfo{person}{Sarah Slobin}.}
  \bibinfo{year}{2014}\natexlab{}.
\newblock \showarticletitle{What If the Data Visualization is Actually People?}
\newblock \bibinfo{journal}{\emph{Source}} (\bibinfo{date}{Apr}
  \bibinfo{year}{2014}).
\newblock
\urldef\tempurl%
\url{https://source.opennews.org/articles/what-if-data-visualization-actually-people/}
\showURL{%
\tempurl}


\bibitem[\protect\citeauthoryear{Talbot, Lee, Kapoor, and Tan}{Talbot
  et~al\mbox{.}}{2009}]%
        {talbot2009ensemblematrix}
\bibfield{author}{\bibinfo{person}{Justin Talbot}, \bibinfo{person}{Bongshin
  Lee}, \bibinfo{person}{Ashish Kapoor}, {and} \bibinfo{person}{Desney~S Tan}.}
  \bibinfo{year}{2009}\natexlab{}.
\newblock \showarticletitle{EnsembleMatrix: interactive visualization to
  support machine learning with multiple classifiers}. In
  \bibinfo{booktitle}{\emph{Proceedings of the SIGCHI Conference on Human
  Factors in Computing Systems}}. ACM, \bibinfo{pages}{1283--1292}.
\newblock


\bibitem[\protect\citeauthoryear{Van~Wijk}{Van~Wijk}{2005}]%
        {van2005value}
\bibfield{author}{\bibinfo{person}{Jarke~J Van~Wijk}.}
  \bibinfo{year}{2005}\natexlab{}.
\newblock \showarticletitle{The value of visualization}. In
  \bibinfo{booktitle}{\emph{Visualization, 2005. VIS 05. IEEE}}. IEEE,
  \bibinfo{pages}{79--86}.
\newblock


\bibitem[\protect\citeauthoryear{Vartak, Rahman, Madden, Parameswaran, and
  Polyzotis}{Vartak et~al\mbox{.}}{2015}]%
        {vartak2015s}
\bibfield{author}{\bibinfo{person}{Manasi Vartak}, \bibinfo{person}{Sajjadur
  Rahman}, \bibinfo{person}{Samuel Madden}, \bibinfo{person}{Aditya
  Parameswaran}, {and} \bibinfo{person}{Neoklis Polyzotis}.}
  \bibinfo{year}{2015}\natexlab{}.
\newblock \showarticletitle{SeeDB: efficient data-driven visualization
  recommendations to support visual analytics}.
\newblock \bibinfo{journal}{\emph{Proceedings of the VLDB Endowment}}
  \bibinfo{volume}{8}, \bibinfo{number}{13} (\bibinfo{year}{2015}),
  \bibinfo{pages}{2182--2193}.
\newblock


\bibitem[\protect\citeauthoryear{Vines, Roesner, and Kohno}{Vines
  et~al\mbox{.}}{2017}]%
        {vines2017exploring}
\bibfield{author}{\bibinfo{person}{Paul Vines}, \bibinfo{person}{Franziska
  Roesner}, {and} \bibinfo{person}{Tadayoshi Kohno}.}
  \bibinfo{year}{2017}\natexlab{}.
\newblock \showarticletitle{Exploring ADINT: Using Ad Targeting for
  Surveillance on a Budget-or-How Alice Can Buy Ads to Track Bob}. In
  \bibinfo{booktitle}{\emph{Proceedings of the 2017 on Workshop on Privacy in
  the Electronic Society}}. ACM, \bibinfo{pages}{153--164}.
\newblock


\bibitem[\protect\citeauthoryear{Wacharamanotham, Subramanian, V{\"o}lkel, and
  Borchers}{Wacharamanotham et~al\mbox{.}}{2015}]%
        {wacharamanotham2015statsplorer}
\bibfield{author}{\bibinfo{person}{Chat Wacharamanotham},
  \bibinfo{person}{Krishna Subramanian}, \bibinfo{person}{Sarah~Theres
  V{\"o}lkel}, {and} \bibinfo{person}{Jan Borchers}.}
  \bibinfo{year}{2015}\natexlab{}.
\newblock \showarticletitle{Statsplorer: Guiding novices in statistical
  analysis}. In \bibinfo{booktitle}{\emph{Proceedings of the 33rd Annual ACM
  Conference on Human Factors in Computing Systems}}. ACM,
  \bibinfo{pages}{2693--2702}.
\newblock


\bibitem[\protect\citeauthoryear{Wachter, Mittelstadt, and Russell}{Wachter
  et~al\mbox{.}}{2017}]%
        {wachter2017counterfactual}
\bibfield{author}{\bibinfo{person}{Sandra Wachter}, \bibinfo{person}{Brent
  Mittelstadt}, {and} \bibinfo{person}{Chris Russell}.}
  \bibinfo{year}{2017}\natexlab{}.
\newblock \showarticletitle{Counterfactual explanations without opening the
  black box: Automated decisions and the GDPR}.
\newblock \bibinfo{journal}{\emph{Harvard Journal of Law \& Technology}}
  \bibinfo{volume}{32}, \bibinfo{number}{2} (\bibinfo{year}{2017}).
\newblock


\bibitem[\protect\citeauthoryear{Wall, Blaha, Franklin, and Endert}{Wall
  et~al\mbox{.}}{2017}]%
        {wall2017warning}
\bibfield{author}{\bibinfo{person}{Emily Wall}, \bibinfo{person}{Leslie~M
  Blaha}, \bibinfo{person}{Lyndsey Franklin}, {and} \bibinfo{person}{Alex
  Endert}.} \bibinfo{year}{2017}\natexlab{}.
\newblock \showarticletitle{Warning, bias may occur: A proposed approach to
  detecting cognitive bias in interactive visual analytics}. In
  \bibinfo{booktitle}{\emph{IEEE Conference on Visual Analytics Science and
  Technology (VAST)}}.
\newblock


\bibitem[\protect\citeauthoryear{Wang, Chen, Chou, Bryan, Guan, Chen, Pan, and
  Ma}{Wang et~al\mbox{.}}{2018}]%
        {wang2018graphprotector}
\bibfield{author}{\bibinfo{person}{Xumeng Wang}, \bibinfo{person}{Wei Chen},
  \bibinfo{person}{Jia-Kai Chou}, \bibinfo{person}{Chris Bryan},
  \bibinfo{person}{Huihua Guan}, \bibinfo{person}{Wenlong Chen},
  \bibinfo{person}{Rusheng Pan}, {and} \bibinfo{person}{Kwan-Liu Ma}.}
  \bibinfo{year}{2018}\natexlab{}.
\newblock \showarticletitle{GraphProtector: A Visual Interface for Employing
  and Assessing Multiple Privacy Preserving Graph Algorithms}.
\newblock \bibinfo{journal}{\emph{IEEE transactions on visualization and
  computer graphics}} (\bibinfo{year}{2018}).
\newblock


\bibitem[\protect\citeauthoryear{Ward}{Ward}{2016}]%
        {ward2016deadly}
\bibfield{author}{\bibinfo{person}{Mark Ward}.}
  \bibinfo{year}{2016}\natexlab{}.
\newblock \bibinfo{booktitle}{\emph{Deadly documents: Technical communication,
  organizational discourse, and the Holocaust: Lessons from the rhetorical work
  of everyday texts}}.
\newblock \bibinfo{publisher}{Routledge}.
\newblock


\bibitem[\protect\citeauthoryear{Williams and Nagel}{Williams and
  Nagel}{1976}]%
        {williams1976moral}
\bibfield{author}{\bibinfo{person}{Bernard~AO Williams} {and}
  \bibinfo{person}{Thomas Nagel}.} \bibinfo{year}{1976}\natexlab{}.
\newblock \showarticletitle{Moral luck}.
\newblock \bibinfo{journal}{\emph{Proceedings of the Aristotelian Society,
  Supplementary Volumes}}  \bibinfo{volume}{50} (\bibinfo{year}{1976}),
  \bibinfo{pages}{115--151}.
\newblock


\bibitem[\protect\citeauthoryear{Wines}{Wines}{2018}]%
        {wines2018}
\bibfield{author}{\bibinfo{person}{Mchael Wines}.}
  \bibinfo{year}{2018}\natexlab{}.
\newblock \showarticletitle{Why Was a Citizenship Question Put on the Census?
  `Bad Faith,' a Judge Suggests}.
\newblock \bibinfo{journal}{\emph{New York Times}} (\bibinfo{date}{Jul}
  \bibinfo{year}{2018}).
\newblock
\urldef\tempurl%
\url{https://www.nytimes.com/2018/07/10/us/citizenship-question-census.html}
\showURL{%
\tempurl}


\bibitem[\protect\citeauthoryear{Wongsuphasawat, Smilkov, Wexler, Wilson,
  Man{\'e}, Fritz, Krishnan, Vi{\'e}gas, and Wattenberg}{Wongsuphasawat
  et~al\mbox{.}}{2018}]%
        {wongsuphasawat2018visualizing}
\bibfield{author}{\bibinfo{person}{Kanit Wongsuphasawat},
  \bibinfo{person}{Daniel Smilkov}, \bibinfo{person}{James Wexler},
  \bibinfo{person}{Jimbo Wilson}, \bibinfo{person}{Dandelion Man{\'e}},
  \bibinfo{person}{Doug Fritz}, \bibinfo{person}{Dilip Krishnan},
  \bibinfo{person}{Fernanda~B Vi{\'e}gas}, {and} \bibinfo{person}{Martin
  Wattenberg}.} \bibinfo{year}{2018}\natexlab{}.
\newblock \showarticletitle{Visualizing dataflow graphs of deep learning models
  in TensorFlow}.
\newblock \bibinfo{journal}{\emph{IEEE transactions on visualization and
  computer graphics}} \bibinfo{volume}{24}, \bibinfo{number}{1}
  (\bibinfo{year}{2018}), \bibinfo{pages}{1--12}.
\newblock


\bibitem[\protect\citeauthoryear{Wood}{Wood}{2010}]%
        {wood2010rethinking}
\bibfield{author}{\bibinfo{person}{Denis Wood}.}
  \bibinfo{year}{2010}\natexlab{}.
\newblock \bibinfo{booktitle}{\emph{Rethinking the power of maps}}.
\newblock \bibinfo{publisher}{Guilford Press}.
\newblock


\bibitem[\protect\citeauthoryear{Wood, Isenberg, Isenberg, Dykes, Boukhelifa,
  and Slingsby}{Wood et~al\mbox{.}}{2012}]%
        {wood2012sketchy}
\bibfield{author}{\bibinfo{person}{Jo Wood}, \bibinfo{person}{Petra Isenberg},
  \bibinfo{person}{Tobias Isenberg}, \bibinfo{person}{Jason Dykes},
  \bibinfo{person}{Nadia Boukhelifa}, {and} \bibinfo{person}{Aidan Slingsby}.}
  \bibinfo{year}{2012}\natexlab{}.
\newblock \showarticletitle{Sketchy rendering for information visualization}.
\newblock \bibinfo{journal}{\emph{IEEE Transactions on Visualization and
  Computer Graphics}} \bibinfo{volume}{18}, \bibinfo{number}{12}
  (\bibinfo{year}{2012}), \bibinfo{pages}{2749--2758}.
\newblock


\bibitem[\protect\citeauthoryear{Wood, Kachkaev, and Dykes}{Wood
  et~al\mbox{.}}{2018}]%
        {wood2018design}
\bibfield{author}{\bibinfo{person}{Jo Wood}, \bibinfo{person}{Alexander
  Kachkaev}, {and} \bibinfo{person}{Jason Dykes}.}
  \bibinfo{year}{2018}\natexlab{}.
\newblock \showarticletitle{Design Exposition with Literate Visualization}.
\newblock \bibinfo{journal}{\emph{IEEE transactions on visualization and
  computer graphics}} (\bibinfo{year}{2018}).
\newblock


\bibitem[\protect\citeauthoryear{Zer-Aviv}{Zer-Aviv}{2015}]%
        {unempathic2015}
\bibfield{author}{\bibinfo{person}{Mushon Zer-Aviv}.}
  \bibinfo{year}{2015}\natexlab{}.
\newblock \bibinfo{title}{DataViz--The UnEmpathetic Art}.
\newblock
  \bibinfo{howpublished}{\url{https://responsibledata.io/2015/10/19/dataviz-the-unempathetic-art/}}.
\newblock


\bibitem[\protect\citeauthoryear{Zgraggen, Zhao, Zeleznik, and Kraska}{Zgraggen
  et~al\mbox{.}}{2018}]%
        {zgraggen2018investigating}
\bibfield{author}{\bibinfo{person}{Emanuel Zgraggen}, \bibinfo{person}{Zheguang
  Zhao}, \bibinfo{person}{Robert Zeleznik}, {and} \bibinfo{person}{Tim
  Kraska}.} \bibinfo{year}{2018}\natexlab{}.
\newblock \showarticletitle{Investigating the Effect of the Multiple
  Comparisons Problem in Visual Analysis}. In
  \bibinfo{booktitle}{\emph{Proceedings of the 2018 CHI Conference on Human
  Factors in Computing Systems}}. ACM, \bibinfo{pages}{479}.
\newblock


\bibitem[\protect\citeauthoryear{Zhu and Porter}{Zhu and Porter}{2002}]%
        {zhu2002automated}
\bibfield{author}{\bibinfo{person}{Donghua Zhu} {and} \bibinfo{person}{Alan~L
  Porter}.} \bibinfo{year}{2002}\natexlab{}.
\newblock \showarticletitle{Automated extraction and visualization of
  information for technological intelligence and forecasting}.
\newblock \bibinfo{journal}{\emph{Technological forecasting and social change}}
  \bibinfo{volume}{69}, \bibinfo{number}{5} (\bibinfo{year}{2002}),
  \bibinfo{pages}{495--506}.
\newblock


\end{thebibliography}

\end{document}